\pdfoutput=1                    
\documentclass[]{aa}
\usepackage{graphicx}
\usepackage{txfonts}
\usepackage{epsfig,graphics,graphicx,bm,amssymb}
\usepackage[normalem]{ulem}

\usepackage[dvipsnames]{xcolor}

\usepackage[]{natbib}
\usepackage[hidelinks]{hyperref}

\begin{document}

\title{
Combined dynamical effects of the bar and spiral arms in a Galaxy model. Application to the solar neighbourhood.  
}

\author{T.\,A. Michtchenko\thanks{e-mail: tatiana.michtchenko@iag.usp.br}
          \and \,J.\,R.\,D. L\'epine\thanks{e-mail: jacques.lepine@iag.usp.br}
          \and \, D.\,A. Barros\thanks{e-mail: douglas.barros@iag.usp.br}
          \and \, R.\,S.\,S. Vieira\thanks{e-mail: rss.vieira@usp.br}
}

\institute{Universidade de S\~ao Paulo, IAG, Rua do Mat\~ao 1226, Cidade Universit\'aria, 05508-090 S\~ao Paulo, Brazil}

\date{}

\abstract
{Observational data indicate that the Milky Way is a barred spiral galaxy. Computation facilities and availability of data from Galactic surveys stimulate the appearance of models of the Galactic structure, however further efforts
are needed to build dynamical models containing both spiral arms and the central bar/bulge.
}
{We expand the study of the stellar dynamics in the Galaxy by adding the bar/bulge component to a model with spiral arms  introduced in one of our previous publications. The model is tested by applying it to the solar neighbourhood, where observational data are more precise.
}
{We model analytically the potential of the Galaxy to derive the force field in its equatorial plane. The model comprises an axisymmetric disc derived from the observed rotation curve, four spiral arms with Gaussian-shaped groove profiles, and a classical elongated/oblate ellipsoidal bar/bulge structure. The parameters describing the bar/bulge are constrained by observations and the stellar dynamics, and their possible limits are determined.
}
{A basic model results in a bar of  2.9\,kpc in length, with a mass of the order of a few 10$^9M_\odot$ (which does not include the axisymmetric part of the bulge, which has a mass of about 10$^{10}M_\odot$). The size and orientation of the bar are also restricted by the position of masers with Very Long Baseline Interferometry (VLBI). The bar's rotation speed is constrained to $\Omega_{\rm bar} < 50$\,km\,s$^{-1}$\,kpc$^{-1}$ taking into account the allowed mass range.
}
{We conclude that our basic model is  compatible with observations and with the dynamical constraints. The model  explains simultaneously the bulk of the main moving groups, associated here with the spiral corotation resonance, and the Hercules stream, associated with several inner high-order spiral resonances; in particular, with the 8/1 resonance. From the dynamical constraints on the bar's angular speed, it is unlikely that the bar's outer Lindblad resonance (OLR) lies near the solar circle; moreover, its proximity would compromise the stability of the local arm structure.
}

\keywords{Galaxies: spiral - Galaxies: kinematics and dynamics - Methods: numerical - Methods: analytical}

\titlerunning{Stellar dynamics in barred spiral galaxies}

\maketitle

\section{Introduction}\label{sec:intro}

The gravitational potential of the Galaxy can be described, with a good approximation, as the sum of the contributions of mass components considered as axisymmetric, that is,  thick disc, thin disc (or only one disc), bulge, and stellar and dark matter halo. The effects of these components add up and reproduce the observed rotation curve \citep[e.g.][]{Allen_Santillan1991,BarrosEtal2016AA}. The total dynamical mass of the Galaxy has been estimated using different methods, and is in the range of 4 to $9\times 10^{11}M_\odot$ \citep{Allen_Santillan1991,GnedinEtalApJL2010,Ablimit_Zhao2017}, with a stellar mass of $\sim5\times 10^{10}M_\odot$ \citep{Bland-Hawthor2016}.

To describe more precisely the stellar orbits in the Galactic disc, one must take into account the two main non-axisymmetric and rotating components: the Galactic bar and the spiral arms. Of special interest in the study of orbits are the existing resonances between the rotation frequency of these components (either bar or spiral arms) and the rotation frequency of the stellar orbits around the Galactic centre.

In previous papers \citep[][hereafter Paper I]{junqueiraEtal2013AA,MichtchenkoEtal2017ApJ}, we developed a new model describing the dynamics of spiral galaxies. The model was applied to the Galaxy, using the observational constraints on the spiral structure and on its pattern rotation speed (L\'epine et al. 2017, hereafter Paper II). It was found that a corotation zone, produced by spiral arms, has established in the solar neighbourhood, with such strong influence on the stellar orbits that the local arm could be explained by this mechanism. Moreover, our investigations of the kinematics in the solar neighbourhood have raised an important issue:  the influence of the bar on the stellar motion.

Many authors have worked on bar models \citep[e.g.][among others]{dehnen2000AJ,pichardoMartosMoreno2004ApJ,BobylevEtal2014, PerezVillegasEtal2017, Portail2017}, and only a few with models of a bar plus spiral arms \cite[][]{quillen2003AJ,antoja2009,antojaEtal2011MNRAS}. This situation sheds doubt on a global view of the stellar dynamics in the disc and makes fruitful comparisons with observations difficult.

Concerning these comparisons, much attention has been given to explaining anomalies in the velocity distribution of stars in the solar neighbourhood, the only region of the Galaxy for which high-quality stellar proper motions are available. The velocity distribution in the $U$--$V$ plane presents a number of structured features, frequently referred to as the moving groups or stellar streams, which have been known for decades \citep[][among others]{Eggen1996,skuljanEtal1999MNRAS,FamaeyEtalAA2005,antojaEtal2011MNRAS}. The main features are the Pleiades-Hyades supercluster and the Sirius cluster. Another feature, the Hercules stream, draws attention due to the high velocity of its stars with respect to the Sun; the stars are lagging behind the local standard of rest (LSR) by about 50 km\,s$^{-1}$ \citep{antoja2008}. \cite{FamaeyEtalAA2005} estimated that about 6\% of the stars in the solar neighbourhood belong to this stream, while \cite{dehnen2000AJ} was the first in giving its dynamical nature.

These anomalies have been tentatively explained by resonances that may occur in the solar neighbourhood. \cite{BensbyEtalApJL2007} argued that the Hercules stream is possibly due to a resonance produced by the bar. \citet{AntojaEtalAA2014} and \citet{MonariEtalMNRAS2017}, among others, have argued that the bar must be fast rotating to explain the Hercules stream by the outer Lindblad resonance (OLR) of the bar near the Sun. It should be noted that, for a given rotation curve of the Galaxy, the Galactic radius of a resonance of the bar depends mainly on its rotation speed, and not on details of the bar model (Paper I).

In the present paper, we add the potential perturbation of the bar to the model of the spiral arms that we have already investigated in Papers I and II. This allows us to compare the magnitude of the perturbation caused by the two components, to observe new interesting features arising from the interaction of the potentials, and to restrict the range of parameters of the bar to be consistent with observational data.

Our model of the gravitational potential of the central region of the Galaxy is composed of two objects: the bar and the bulge. This separation was usual in the past \citep[e.g.][Chapter 1]{Mihalas_Binney1981}, but is not always made at present. The bulge is the massive spheroidal component at the centre of the Galaxy, but, due to its symmetry, it does not participate in the formation of resonances in the Galactic equatorial plane. The bar is the less massive, elongated structure that is responsible for resonances and interactions with the spiral arms. We do not focus on the stellar orbits internal to the bar, which maintain its elongated shape, but only on the dynamical interactions taking place outside the physical volume of the bar. Looking for dynamical constraints, we vary the parameters of the bar, such as its total mass, size, current orientation, flattening and rotation velocity, and discuss their acceptable ranges and positions of resonances.

Among the conclusions that we come to, we argue that it is possible that the bar rotates with the same velocity as the spiral arms.
The main moving groups in the solar neighbourhood may be associated with the spiral corotation zone, and the Hercules stream is best explained by the 8/1 resonance of the spiral structure. Moreover, we show that it is unlikely that the bar's OLR lies near the solar radius, since this position could produce strong instabilities inside the local arm region due to the overlap between the spiral corotation resonance and the bar's OLR. We also suggest an explanation for the nature of the “long bar” with dimensions of about 4.5 kpc \citep{Lopez_CorredoiraEtal2007}, which contrasts with the more generally accepted “short bar” \citep[e.g.][]{BobylevEtal2014,Wegg2015}.

The structure of the paper is as follows. In Sect.\,\ref{sec:model}, we present the model with all theoretical assumptions and justifications for the adopted parameters based on observations. In Sect.\,\ref{sec:topology}, we investigate the topology of the perturbation potential and analyse the corotation zones produced by the spiral and bar/bulge resonances. In Sect.\,\ref{sec:dyn-map}, dynamical maps on the $R$-–$V_\theta$ plane and constraints on the bar's mass are presented; the constraints on other parameters of the bar are further discussed in Sect.\,\ref{sec:map-param}, and the effects of the bar's rotation speed in Sect.\,\ref{sec:bar-speed}. Additional support to the model coming from observations, such as the distribution of maser sources, the main moving groups, and the Hercules stream, are explained in Sect.\,\ref{sec:evidence}, and the conclusions are presented in Sect.\,\ref{sec:conclus}.

\begin{table}
\caption{Basic values of the physical and geometrical parameters adopted in the model.}
\label{tab:1}       
\begin{tabular}{lclc}
\hline 
\hline
 & & &  \\
Parameter & Symbol & Basic value & Units  \\
\hline
\hline
Sun's galactic radius          & $R_\odot$             & 8.0        & kpc  \\
LSR velocity          & $V_0$                 & 230        & km s$^{-1}$  \\
\hline
Spiral arms  & &      &   \\
\hline
Number of arms        & m                     & 4          & -   \\
Spiral pattern speed  & $\Omega_{\rm sp}$     & 28.5       & km\,s$^{-1}$\,kpc$^{-1}$\\
Pitch angle           & i                     & -14$^\circ$& -   \\
Arm width             & $\sigma \sin {\rm i}$ & 1.94       & kpc \\
Scale length          & $\varepsilon_s^{-1}$  & 4.0        & kpc  \\
Spiral amplitude      & $\zeta_0$             & 200.0      & km$^2$\,s$^{-2}$\,kpc$^{-1}$\\
Reference radius      & $R_i$                 & 8.0        & kpc\\
\hline
Central bar  & &      &   \\
\hline
Bar's mass            & $M_{\rm bar}$      & 1$\times 10^{9}$ &$M_\odot$\\
Bar's radius          & $R_{\rm bar}$      & 2.9              &kpc\\
Equatorial flattening & $f_{\rm bar}$      & 0.7              & - \\
Bar's initial phase   & $\gamma^0_{\rm bar}$ & 67$^\circ.5$     & - \\
Bar's pattern speed   & $\Omega_{\rm bar}$ & 28.5             & km\,s$^{-1}$\,kpc$^{-1}$\\
\hline
Central bulge  & &      &   \\
\hline
Bulge's mass          & $M_{\rm bulge}$ & 1$\times 10^{10}$ &$M_\odot$\\
Bulge's radius        & $R_{\rm bulge}$ & 1.0              &kpc\\
Polar flattening      & $f_{\rm bulge}$ & 0.2              & - \\
\hline
\end{tabular}
\end{table}

\begin{table}
\caption{Tested limits of the bar's parameters.}
\label{tab:1-1}       
\begin{tabular}{lccc}
\hline 
\hline
 & & &  \\
Parameter & Symbol & Tested limits & Units  \\
\hline
\hline
Bar's mass            & $M_{\rm bar}$      &$10^{8}$--$10^{11}$  &$M_\odot$\\
Bar's radius          & $R_{\rm bar}$      & 1 -- 6              &kpc\\
Equatorial flattening & $f_{\rm bar}$      & 0.1 -- 0.9          & - \\
Bar's initial phase   & $\gamma^0_{\rm bar}$ & 0 -- 180$^\circ$    & - \\
Bar's pattern speed   & $\Omega_{\rm bar}$ & 15 -- 70            & km\,s$^{-1}$\,kpc$^{-1}$\\
\hline
\end{tabular}
\end{table}


\section{Model}\label{sec:model}

Our model considers the Galaxy composed of three main components interacting gravitationally: the axisymmetric disc, the four-arm spiral structure and the central bar/bulge component\footnote{The stellar and dark-matter halos are naturally included in the observed rotation curve (see Sect.\,\ref{sec:rot-curve})}.  The two-degrees-of-freedom Hamiltonian, which describes the stellar motion in the equatorial plane of the Galaxy, can be written in the rotating frame as
\begin{equation}\label{eq:eq1}
{\mathcal H}(R,\varphi,p_r,L_z, t)= {\mathcal H_{0}}(R,p_r,L_z) + {\mathcal H}_1(R,\varphi, t),
\end{equation}
with ${\mathcal H_{0}}$ and ${\mathcal H}_1$ being the unperturbed and perturbation contributions, respectively. The perturbation is composed of three parts as
\begin{equation}\label{eq:eq1-p}
{\mathcal H}_1(R,\varphi, t)= \Phi_{\rm{sp}}+\Phi_{\rm{bar}}+\Phi_{\rm{bulge}},
\end{equation}
where $\Phi_{\rm{sp}}$, $\Phi_{\rm{bar}}$, and $\Phi_{\rm{bulge}}$ are perturbations due to spiral arms, bar, and bulge, respectively. 
The polar coordinates (the Galactocentric radius and azimuthal distance, $R$ and $\varphi$, respectively) are defined in the rotating frame, with the angular speed $\Omega_p$.  The canonical momenta $p_r$ and \mbox{$L_z$} are the linear and angular momenta per unit mass, respectively.


\subsection{Axisymmetric potential and rotation curve}\label{sec:rot-curve}

The one-degree-of-freedom unperturbed part of the Hamiltonian (\ref{eq:eq1}) is given by Jacobi's integral as
\begin{equation}\label{eq:H0}
{\mathcal H_{0}}(R,p_r,L_z)= \frac{1}{2}\left[ p_r^2 + \frac{L_z^2}{R^2}\right] - \Omega_p L_z + \Phi_0(R),
\end{equation}
where the first term defines the kinetic energy of a star and the second is a gyroscopic term due to the rotating reference frame, with pattern speed $\Omega_p$. In this work, we generally assume the spiral arms' speed $\Omega_{\rm sp}$ as the pattern speed, $\Omega_p=\Omega_{\rm sp}$, and, only in the case when only perturbations due to the central bar/bulge  are considered, do we assume the bar's rotation speed as a pattern speed.

The last term in Eq. (\ref{eq:H0}), $\Phi_0(R)$, is the Galactic axisymmetric potential defined by the rotation curve $V_{\rm rot}(R)$ via the relation
\begin{equation}
\partial\Phi_0/\partial R = V^2_{\rm rot}(R)/R\,.
\label{eq:axisymmetric}
\end{equation}
We adopt a realistic rotation-curve model of the Milky Way  based on published observational data \citep{clemens1985ApJ,fichBlitzStark1989ApJ,reidEtal2014ApJ}
, which we fit by the sum of two exponentials in the form (see 
Paper I for details of the data used and the fitting procedure)
\begin{eqnarray}
 V_{\rm rot}(R) &=& 298.9\,\exp\left(-\frac{R}{4.55}-\frac{0.034}{R}\right)  \nonumber\\
 & & + 219.3\,\exp\left[-\frac{R}{1314.4}-\left(\frac{3.57}{R}\right)^2\right] ,
 \label{eq:Vrot}
\end{eqnarray}
\noindent
with the factors multiplying the exponentials given in units of kilometers per second and the factors in the arguments of the exponentials given in kiloparsecs. In order to obtain the axisymmetric gravitational potential $\Phi_0(R)$, we solve Eq.~(\ref{eq:axisymmetric}) numerically, using the trapezium rule with adaptive step  and adopting the ``numerical infinity'' condition $\Phi_{0}(1000\,{\rm kpc})=0$.

It is worth noting that our approximation of $\Phi_0(R)$ requires no assumptions on which components of the Galaxy (stellar or gaseous matter, dark matter, etc.) are effectively contributing to the axisymmetric potential at each radius. However, it must be kept in mind that radial components (i.e. the components which are independent from the azimuthal angle $\varphi$)  of the central bar and bulge potentials are already accounted for in the expression in Eq.(\ref{eq:axisymmetric}).
\begin{figure}
\begin{center}
\epsfig{figure=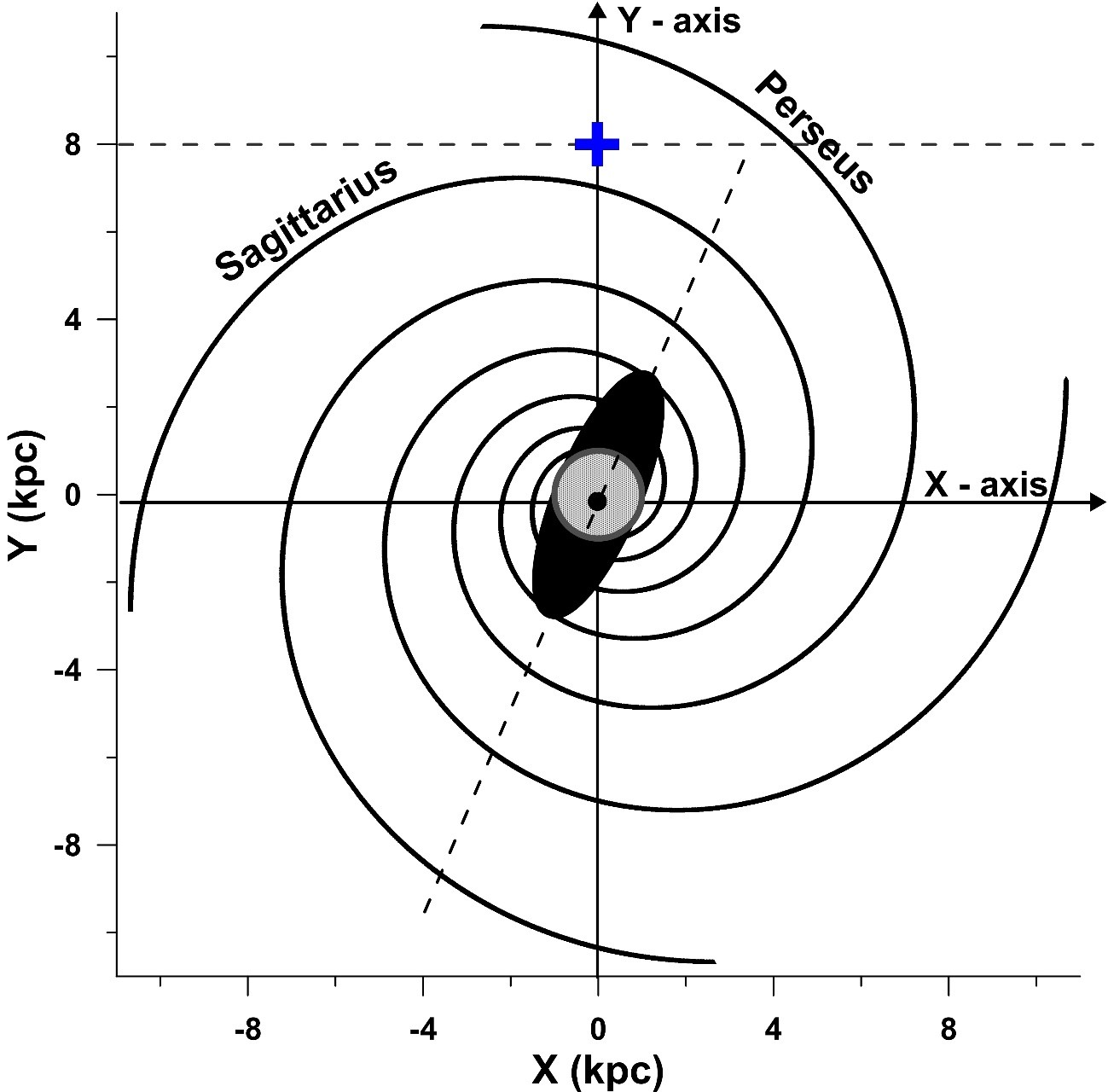,width=0.99\columnwidth ,angle=0}
\caption{Schematic view of the equatorial $X$--$Y$ plane of a four-arm spiral barred galaxy in the adopted reference frame. The main perturbation components, such as the spiral arms, the central bar, and the bulge, were calculated using the parameter's values from Table\,\ref{tab:1}, except for the pitch angle $i=+14^\circ$. The position  of the Sun is  shown by a blue cross. The bar's phase with respect to the reference direction ($X$--axis) is $67^\circ.5$ (or $22^\circ.5$ with respect to the Sun).  Close to the Sun, Sagittarius and Perseus arms are identified. }
\label{fig:sketch}
\end{center}
\end{figure}


\subsection{Spiral arms' potential}\label{sec:model-sp}

The spiral arms' two-dimensional (2D) potential is introduced as a logarithmic perturbation to the axisymmetric potential $\Phi_0(R)$. We adopt the Gaussian-shaped azimuthal groove profile for the spiral potential as described in \cite{junqueiraEtal2013AA}:
\begin{equation}
\label{eq:H1gaussian}
 \Phi_{\rm sp}(R,\varphi) = -\zeta_0\,R\,e^{-\frac{R^2}{\sigma^2}[1-\cos(m\varphi-f_m(R))]-\varepsilon_s R},
\end{equation}
where $m=4$ is the number of arms and $\varphi$ is the azimuthal angle in the frame rotating with angular velocity $\Omega_p=\Omega_{\rm sp}$.
The shape function $f_m(R)$ is given by
\begin{equation}\label{eq:eq6}
f_m(R) = \frac{m}{\tan(i)}\ln{(R/R_i)}+ \gamma\, ,
\end{equation}
where $i$ is the spiral pitch angle; $R_i$ is a reference radius and $\gamma$ is an arbitrary phase angle, whose values define the orientation of the spirals in the chosen reference frame. The values of the parameters in Eqs.\,(\ref{eq:H1gaussian}) and (\ref{eq:eq6}) adopted in this work, and their physical meanings are given in Table\,\ref{tab:1} (the detailed discussion on this choice can be found in Papers I and II).

Figure\,\ref{fig:sketch} shows, by black curves, the loci of four arms on the ($X=R\,\cos\varphi$,\,$Y=R\,\sin\varphi$)--plane, which were obtained as azimuthal minima of the potential (Eq. \ref{eq:H1gaussian}) in the reference frame defined as follows. The origin of the reference frame lies at the Galactic centre, while the equatorial plane of the Galaxy is defined as a reference plane. The axis $X$ ($\varphi=0$) is fixed in such a way that the Sun's azimuthal coordinate  is $\varphi = 90^\circ$, placing the Sun on the $Y$-axis at $R=8.0$\,kpc  (blue cross in Fig.\,\ref{fig:sketch}). The orientation of the spiral arms on the $X$--$Y$ plane is fixed by the value of the free parameter $\gamma$ in Eq. (\ref{eq:eq6}). We choose $\gamma$-value such that the Sun (located at $R=8.0$\,kpc and $\varphi=90^\circ$) is 1\,kpc from  the Sagittarius arm locus. Thus we obtain $\gamma=237^\circ.25$ for the spirals parameters from Table\,\ref{tab:1}.

It is worth noting here the choice of the sign of the pitch angle $i$. The conventional maps of the spiral structure of the Milky Way (e.g. \citealt[among others]{georgelinGeorgelin1976AA, drimmelSpergel2001ApJ, russeil2003AA, vallee2013IJAA, houHan2014AA})
present the Galactic rotation in the clockwise direction from the viewpoint of an observer located towards the direction of the North Galactic Pole. In order to follow this convention, the sign of the pitch angle will be chosen positive for the presentation of the results obtained.


\subsection{Central bar and bulge potentials}\label{sec:model-bar}

We adopt simple models for the Galactic central bar and bulge, which draw the bar and bulge as homogeneous elongated and oblate ellipsoids\footnote{The elongated and oblate ellipsoids are classical figures of equilibrium known as the Jeans and MacLaurin ellipsoids, respectively. Both are satisfactory first-order approximations of more complex figures of the bar and bulge \citep{chandrasekhar1987}.},  respectively. Denoting the main semi-axes of the ellipsoids as $a$, $b,$ and $c$, we have $a>b=c$, for the bar, and $a=b>c$, for the bulge. Then, we introduce the equatorial and polar flattenings of the bar and the bulge, respectively, as
\begin{equation}\label{eq:flat}
f_{\rm bar} = \frac{a-b}{a} \hspace{0.5cm} \mbox{and}  \hspace{0.5cm} f_{\rm bulge} = \frac{a-c}{a}.
\end{equation}
They are free parameters of our model and the range of their possible values is analysed in this paper.

According to classical theories of potential, the gravitational potential $U$ generated by a homogeneous ellipsoid at a point $P$, with coordinates $x^*$, $y^*$ and $z^*$ in the reference frame with the origin at the centre of mass of the ellipsoid and the axes aligned to its main semi-axes, can be written in a generic form as \citep{Duboshin1968}
\begin{equation}\label{eq:generic-1}
U(P) = -\frac{3}{2}G\,M\big[U_0(\zeta)+U_1(\zeta)\,x^{*2}+U_2(\zeta)\,y^{*2}+U_3(\zeta)\,z^{*2}\big],
\end{equation}
where $G$ is the gravitational constant, $M$ is the total mass of the ellipsoid and the coefficients $U_0(\zeta)$, $U_1(\zeta)$, $U_2(\zeta)$ and $U_3(\zeta)$ are complex functions of the argument $\zeta$ (Eq.\,\ref{zeta}). In the case of elongated  and oblate ellipsoids, the coefficients can be written explicitly in elementary functions (see Appendix\,\ref{app-A}).

It is worth noting that the axisymmetric contribution of the potential $U(P)$  has already been taken into account in the axisymmetric potential $\Phi_0$ in Eq.\,\ref{eq:axisymmetric} via the observable rotation curve (Eq. \ref{eq:Vrot}), as discussed above. Therefore, we subtract a spherical approximation of this contribution from the total $U(P)$ at the point $P(x^*,y^*,z^*)$, using  Eq.\,\ref{eq:sphere} in Appendix A. It would  be interesting to point out similarities and differences between our bar's model and that introduced in \cite{dehnen2000AJ}. Indeed, a complex mass structure can be formally presented by a multipole expansion: Denhen's model is limited to a low-order quadrupole term in this series, while our model, extended to higher orders, is more precise. Moreover, the total mass of the bar appears explicitly in our model, which allows us to investigate its possible limits. Finally, our model of the bar can be extended to three dimensions immediately.

In this paper, we study the motion of a star in the equatorial Galactic plane, where the star's position is given by the polar coordinates $R$ and $\varphi$ (and $z=0$). Due to the symmetry of the bulge, its potential is axisymmetric in this plane, that is, $U_1(\zeta)=U_2(\zeta)$ in Eq.\,(\ref{eq:generic-1}), and the polar coordinates $R$ and $\varphi$ can be promptly transformed to the rectangular coordinates \mbox{$x^*=R\cos\varphi$} and \mbox{$y^*=R\sin\varphi$}. In the case of the bar, the transformation to the bar's reference frame is
\begin{equation}\label{eq:gamma-bar}
 x^*=R\cos{[\varphi-\gamma_{\rm bar}(t)]},\hspace{.3cm}y^*=R\sin{[\varphi-\gamma_{\rm{bar}}(t)]}.
\end{equation}
Here $\gamma_{\rm{bar}}(t)$ is the azimuthal phase of the bar, whose motion is assumed to be a uniform rotation with the angular speed $\Omega_{\rm{bar}}$ with respect to an inertial frame.

The initial orientation of the central bar $\gamma^0_{\rm bar}$ with respect to the $X$--axis can be defined assuming that two of the four spiral arms are initially connected with the extremities of the bar\footnote{This is an arbitrary assumption based on a visual inspection of several images of barred galaxies. 
We consider several different initial orientations of the bar in Sect.\,\ref{sec:map-param}.}. This condition implies that the initial phase of the bar is given as
\begin{equation}\label{eq:barsize}
\gamma^0_{\rm bar}= f_m(R_{\rm{bar}})/m,
\end{equation}
where $R_{\rm{bar}}$ is the semi-major axis (or radius) of the central bar, $m=4$ and $f_m$ is the shape function given in Eq.\,(\ref{eq:eq6}).
From observations, the current bar's phase with respect to the Sun varies in the range from 10$^\circ$  to 30$^\circ$ (i.e., from 60$^\circ$ to 80$^\circ$ with respect to the $X$--axis) \citep{BobylevEtal2014}; therefore, using the above expression, we can assume the bar's radius to be in the range from 2.78\,kpc to 3.05\,kpc (see details in Sect.\,\ref{sec:model-mass-size}).

Figure\,\ref{fig:sketch} shows the projection of the central bar on the equatorial plane as a very eccentric ellipse, with the semi-major axis  $R_{\rm{bar}}=2.9$\,kpc and the initial phase of $\gamma^0_{\rm{bar}}=67^\circ.5$, with respect to the $X$--axis. This value of $\gamma^0_{\rm{bar}}$ places the bar's semi-major axis at an angle of $22^\circ.5$ with respect to the direction Sun--Galactic centre. The eccentricity of the ellipse is defined by the bar's flattening, whose starting value is  chosen as $f_{\rm bar}=0.7$ (see Table\,\ref{tab:1}).

Finally, Fig.\,\ref{fig:sketch} also shows the projection of the central bulge on the galactic equatorial plane as a circle. The radius of the circle is assumed to be 1\,kpc and the polar flattening $f_{\rm bulge}=0.2$. The potential of the oblate ellipsoid that describes the central bulge introduces no asymmetric perturbations to the star's motion on the equatorial plane. On the other hand, the axisymmetric perturbations are already included in the unperturbed potential $\Phi_0(R)$ via the rotation curve, as discussed in Sect.\,\ref{sec:rot-curve}. Therefore, the bulge's contribution, $\Phi_{\rm{bulge}}(R)$, in the Hamiltonian (\ref{eq:eq1-p}), can be disregarded in the case when the stellar motion is confined to the equatorial plane. Despite this fact, in this paper, we consider the term $\Phi_{\rm bulge}(R)$ always together with the term $\Phi_{\rm bar}(R,\varphi, t)$, referring to them as a bar/bulge perturbation.

\begin{figure}
\begin{center}
\epsfig{figure=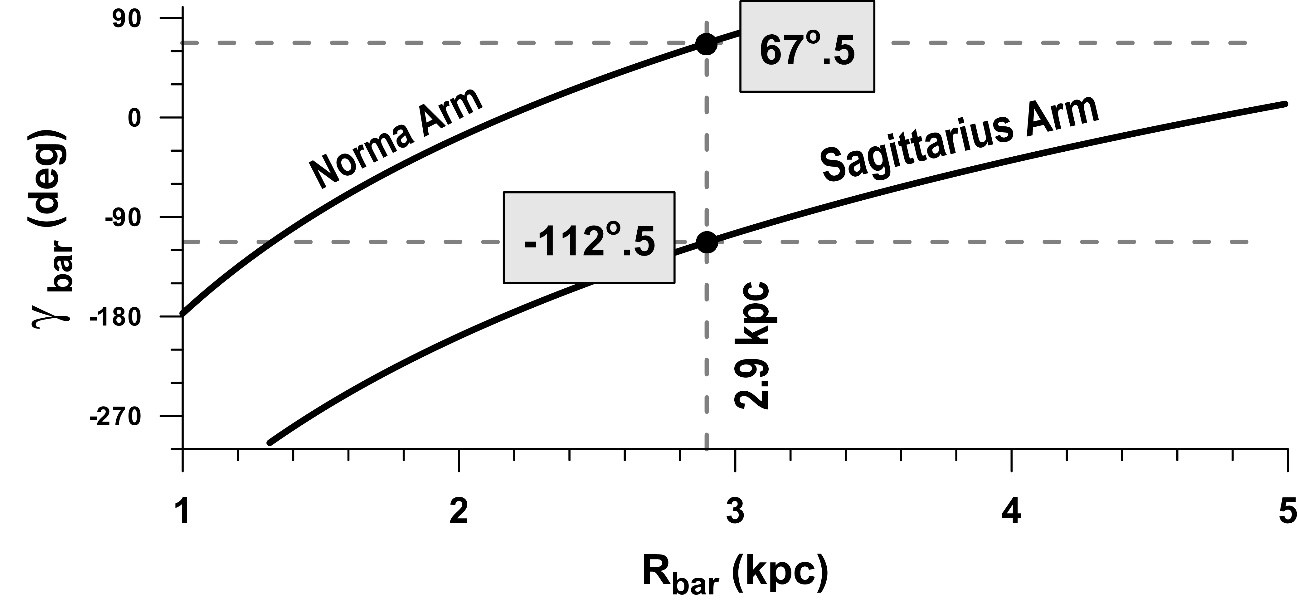,width=0.99\columnwidth ,angle=0}
\caption{Relation between the bar's size $R_{\rm bar}$ and orientation $\gamma_{\rm bar}$ defined by the Sagittarius and Norma Arms (continuous solid curves), which touch the extremes of the bar, at a given value of $R_{\rm bar}$. For $R_{\rm bar}=2.9$\,kpc, the Sagittarius arm reaches the far extremity of the bar at $\gamma_{\rm bar}=-112^\circ.5$, while the Norma arm reaches the near extremity of the bar at $\gamma_{\rm bar}=67^\circ.5$. This last value is supported by observations.
}
\label{fig:barradius}
\end{center}
\end{figure}

\subsection{Choice of the mass of the central bar and bulge, and the bar's rotation speed and size}\label{sec:model-mass}

The application of the model of the bar/bulge requires the knowledge of the parameters of this structure. From these, we paid special attention to the choice of the total mass, rotation speed, and size of the bar. Indeed, as is shown below, the values of these parameters are crucial for the stability of solar motion inside the corotation zone, the stable corotation island encompassing the Sun (see Paper II).

\subsubsection{Setting the bar/bulge mass}\label{sec:model-mass-mass}

The approach adopted in this paper is to consider the bar and the bulge separately. This is for convenience but is also based on observational evidence. In a paper analyzing the Spitzer Survey of Stellar Structure in Galaxies (S4G), Salo et al. (2015) decomposed the brightness profiles of discs and bulge-bar regions of a large number of galaxies. Furthermore, Meidt et al. (2014) investigated the conversion factor of light to mass, and concluded that it is a good approximation to consider the same factor for distinct stellar populations observed at 3.6 microns. Therefore, the brightness distributions obtained by  Salo et al. (2015) can be directly interpreted as being equivalent to mass  distributions. In many cases, the two components, bar and bulge, were distinguished and fitted in Salo et al. (2015). The brightness profiles of the bulges, obtained by these authors, closely resemble the density profile of the bulge of our Galaxy, which can be derived from the rotation curve.

The peak of the rotation curve at about 300\,pc from the centre is attributed to the bulge \citep{Lepine_Leroy2000}. \cite{sofueNakanishi2016PASJ}, from a study of spiral galaxies similar to the Milky Way, found  an average value of 2.3$\times 10^{10} M_\odot$ for the mass of their bulges. In a specific study of the rotation curve of the Galaxy, \cite{sofue2013PSSS} found 1.80$\times 10^{10} M_\odot$  for
the bulge. The two works on our Galaxy that we have mentioned are based on fitting the rotation curve, but do not include the bar among the components, since the contribution of the bar is not clearly apparent. Therefore, we infer that the mass of the bar is included in their estimated values of the bulge mass, and, in addition, that the mass of the bar is smaller than that of the bulge. Indeed, in galaxies which resemble the Milky Way, such as NGC 5101 SB and IC4901 SBbc,  the bar/bulge ratios are  0.76 and 0.06, respectively, that is, the mass of the bar is smaller than that of the bulge. In a recent model, considering the density of red clump stars from different surveys, \citet{Portail2017}  concluded that the stellar mass of the bar/bulge is $1.88\times 10^{10}M_\odot$, of which $1.34\times 10^{10}M_\odot$ is located in the bulge and $5.4\times 10^{9}M_\odot$ in the ``long'' bar.

The aim of the above discussion was to justify that, in terms of orders of magnitude, the values of the total masses of the bar and the bulge of the Galaxy, adopted as $\sim10^{9}M_\odot$ and \mbox{$\sim10^{10}M_\odot$}, respectively, are good initial approximations. It should be stressed that our purpose is to explore orders of magnitude of the mass of the bar, and not to perform fine tuning, for the moment. The mass values of the bulge and of the bar inserted in Table\,\ref{tab:1} should be considered as a first step, or basic model, from which we start the exploration of the range of parameters, as described in the following sections.   Due to the nature of the present work, which focusses on non-axisymmetric components, we are only able to find dynamical constraints on the mass of the bar, but not on that of bulge.
\begin{figure}
\begin{center}
\epsfig{figure=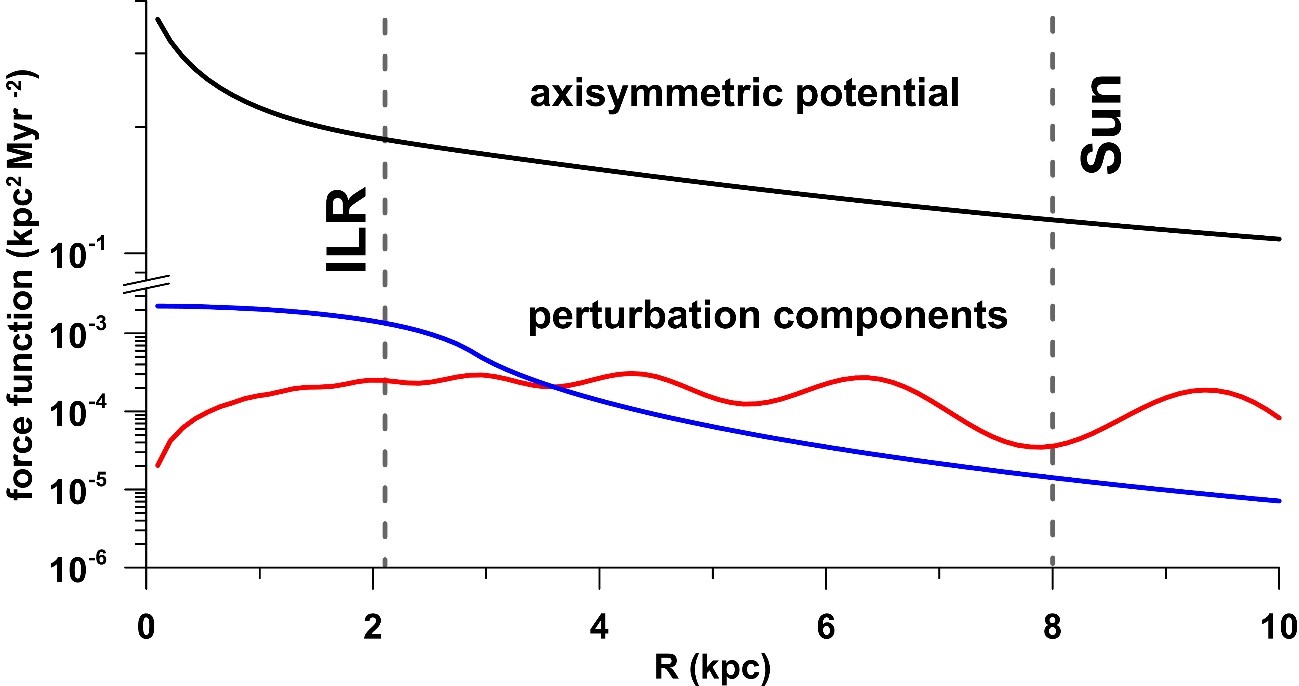,width=0.99\columnwidth ,angle=0}
\caption{Top: Axisymmetric potential $\Phi_0(R)$ from Eq.\,(\ref{eq:H0}) (units are $M_\odot$, kpc, and Myr). Positions of the Sun and the ILR are indicated by the vertical lines at 8\,kpc and 2.13\,kpc, respectively. Bottom: The equatorial non-axisymmetric components of the perturbation potential: the spiral arms perturbation, $\Phi_{\rm sp}(R,\varphi)$ (red curve), and the bar perturbation, $\Phi_{\rm{bar}}(R,\varphi)$ (blue curve). The calculations were done in direction of the bar's major axis, with the phase $\varphi=67^\circ.5$, and with the parameters taken from Table\,\ref{tab:1}. We note that all components are plotted with positive sign, in logarithmic scale; in this case, they are referred to as force functions.}
\label{fig:forcefunction}
\end{center}
\end{figure}

\subsubsection{Setting the bar's rotation speed}\label{sec:model-mass-omega}

The rotation speed of the bar is another parameter which seems to be polemical. Indeed, there is still no observational evidence which could be used to constrain its value, but only indirect measurements through the dynamical effects produced by the central bar on stellar orbits. For this reason, different values for the  bar's speed have been assumed in the literature, and  the kinematic observables are fitted for a range of $\Omega_{\rm bar}$. A recently obtained good fit gives $\Omega_{\rm bar}=39\pm 3.5$\,km\,s$^{-1}$\,kpc$^{-1}$ \citep{Portail2017}. The stellar kinematics in the close neighbourhood of the Sun and in particular the bimodality in the $U$--$V$--velocity distribution (the Hercules stream) have also been used to evaluate the bar's rotation speed, but, depending on adopted hypotheses, this approach provides very different values \citep{dehnen2000AJ,AntojaEtalAA2014,Bienayme2017}. Moreover, these authors usually neglect the role of the spiral arms in the local kinematics, which, in our view, is not justified.

N-body simulations seem to be able to partially produce the bar/bulge structure in the Milky Way and to estimate its rotational speed \citep[e.g.][]{MillerSmith1979, CombesElmegreen93}.  However, due to the high complexity of the processes involved in the study and a plethora of unknown parameters,  any numerical work is limited to having an illustrative, rather than conclusive, character, even if flawless.

As we show in this paper, the rotation speed of the bar is a little restricted by dynamical considerations; it could be any value over a wide range. We made the initial choice adopting the same pattern value of the spiral arms, that is $\Omega_{\rm bar}=\Omega_{\rm sp}=28.5$\,km\,s$^{-1}$\,kpc$^{-1}$, simply because in this case the Hamiltonian (\ref{eq:eq1}) is independent of time.
It should be stressed that this initial assumption can be justified in  a number of ways. The visual inspection of many images of barred galaxies show that, in general, two spiral arms seem to start at the extremities of the bar. If the bar were rotating at a different speed, this connection would suffer a rupture. In particular, in our Galaxy, no such evidence of a rupture has been reported. On the contrary, the known spiral arms, as described by the parameters in Paper II, seem to match the bar extremities well. Among other arguments in favour of a common rotation speed of the bar and of the spiral arms, is that if the bar were rotating faster, tidal effects would tend to slow it down until it reaches synchronization. Interacting galaxies provide good examples of such tidal effects. However, studies of this process are rare \citep{Lokas2016ApJ}.

Furthermore, there are models according to which the spiral arms are formed by the bar, resulting in a single rotating structure \citep{SormaniBinney2015}, and others in which the contrary happens.
From a cosmogonical point of view, the bar and the spiral arms could be considered as a single structure, in which, for instance, two components are synchronized by tidal torques. The (near-)alignment of the main axes of the bar with the centres of the four corotation islands (in particular, with the $L_4$ centre) may be an indication of this configuration.  In this way, (as
initial test value) we can adopt  the rotation speed of the bar as being equal to the speed of the spirals, 28.5\,km\,s$^{-1}$\,kpc$^{-1}$, which was successfully used for the spiral pattern in Paper II. This parameter is tested in the range from 15 to 70\,km\,s$^{-1}$\,kpc$^{-1}$ and the constraints on its value are derived (see Sect.\,\ref{sec:bar-speed}). Finally, we claim that the parameters of the bar/bulge should be derived by modelling the origin of this structure (instead of assuming a priori hypotheses or performing numerical simulations) and compared to observations, which could support this model.

\subsubsection{Setting the bar's size}\label{sec:model-mass-size}

Assuming that two of the four spiral arms are initially connected with the extremities of the bar and setting the initial orientation of the bar with respect to the $X$--axis, we restrict the size of the bar. Indeed, the phase $\gamma_{\rm bar}$ is defined by the radius of the central bar through the relation in Eq. (\ref{eq:barsize}) and Fig.\,\ref{fig:barradius} shows $\gamma_{\rm bar}$ as a function of the bar's radius $R_{\rm{bar}}$. The radius $R_{\rm{bar}}=2.9$\,kpc provides the phase of the far extremity as $\gamma_{\rm bar}=67^\circ.5$, that places the bar's major axis at $22^\circ.5$ with respect to the Sun.
\begin{figure}
\begin{center}
\epsfig{figure=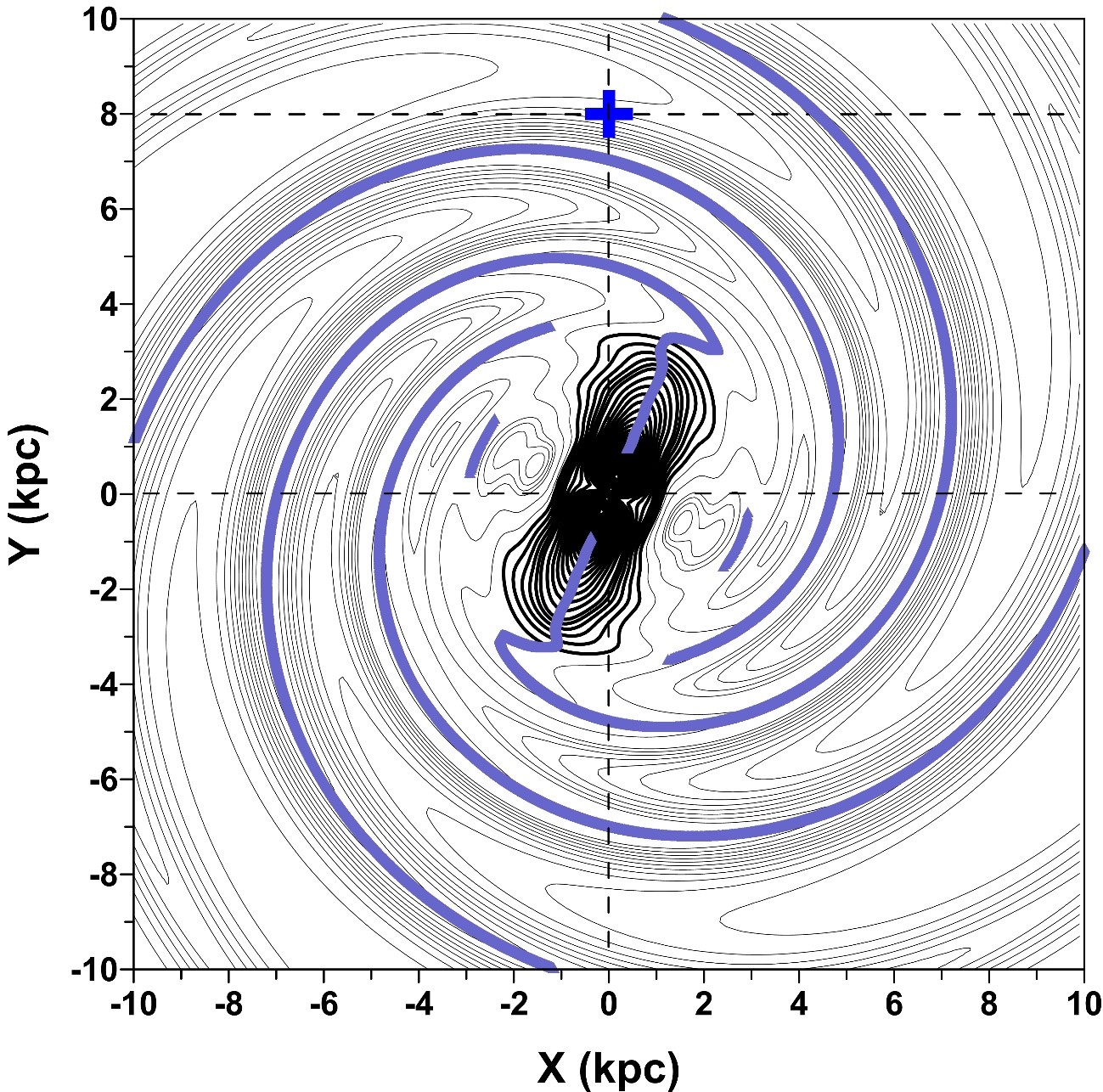,width=0.99\columnwidth ,angle=0}
\caption{Energy levels of the perturbation potential ${\mathcal H}_1(R,\varphi)$ on the representative $X$--$Y$--plane. The levels for which the bar's perturbations are dominating are emphasised, showing the zone of influence of the bar. The position of the Sun is shown by a blue cross.  The blue spiral curves are the azimuthal minima of the potential ${\mathcal H}_1(R,\varphi)$: outside the zone of influence of the bar; they correspond to the loci of the main spiral arms in our model. In calculations, we used the parameter's values from Table\,\ref{tab:1}. }
\label{fig:U1}
\end{center}
\end{figure}

A number of authors believe that there is a connection between the size of the bar and its rotation speed that imposes severe restrictions; for example, that the length of the bar should be close to its corotation radius \citep{Contopoulos1980}. However, this statement is not valid for late-type galaxies like the Milky Way \citep{ElmegreenElmegreen1989ApJ,CombesElmegreen93}. In the case of the Galaxy, there is no requirement to have a bar of 2.9\,kpc with corotation radius close to the solar radius.


\section{Topology of the perturbation potential and corotation zones }\label{sec:topology}

If $\Omega_{\rm bar} = \Omega_{p}$, the Hamiltonian given in Eq.\,(\ref{eq:eq1}) is independent of time in the common rotating frame. The perturbation potential ${\mathcal H}_1(R,\varphi)$ is visualised in Fig.\,\ref{fig:forcefunction}, where its equatorial non-axisymmetric components, $\Phi_{\rm sp}(R,\varphi)$ and $\Phi_{\rm{bar}}(R,\varphi),$ are shown as functions of the Galactic radius. For the sake of comparison, we also plot the axisymmetric unperturbed potential at the top of the figure. We note that the gravitational potentials are plotted with opposite (positive) sign; to avoid possible misunderstanding, we refer hereafter to the gravitational potential with positive sign as \emph{a force function}. The  larger magnitude of the force function will manifest itself as a stronger perturbation on the stellar motion.

The two components of the perturbation shown in Fig.\,\ref{fig:forcefunction} were calculated with the basic set of parameters from Table\,\ref{tab:1} and in the direction of the bar's major axis, with $\varphi=67^\circ.5$. At radial distances of the Sun (vertical line at $R=8$\,kpc), their magnitudes are at least 10$^{4}$ times smaller, when compared to the magnitude of the axisymmetric term $\Phi_{0}(R)$, shown at the top of the panel. This fact clearly characterises ${\mathcal H}_1$ as a small perturbation to $\Phi_{0}$ in the solar neighbourhood. Moreover, the magnitude of the force function of the bar (blue curve) in this region is only a quarter of that of the spiral arms' force function (red curve). We note that the Sun's position is close to one local minimum of the spiral force function (red curve): for the  pattern speed adopted as $\Omega_p=28.5$\,km\,s$^{-1}$\,kpc$^{-1}$, this minimum lies in the proximity of the corotation radius. We should mention that we have not applied any cutoff radius to the arms at their inner boundary. The spiral structure is usually believed to start at the inner Lindblad resonance (ILR), whose position is shown by a vertical dashed line at $R=2.13$\,kpc in Fig.\,\ref{fig:forcefunction}. However, since this resonance is located inside the bar, in a region where the effect of the arms is already negligible (with magnitude of perturbation at least one order smaller than that of the bar), this option has no effect on the results.

\begin{figure}
\begin{center}
\epsfig{figure=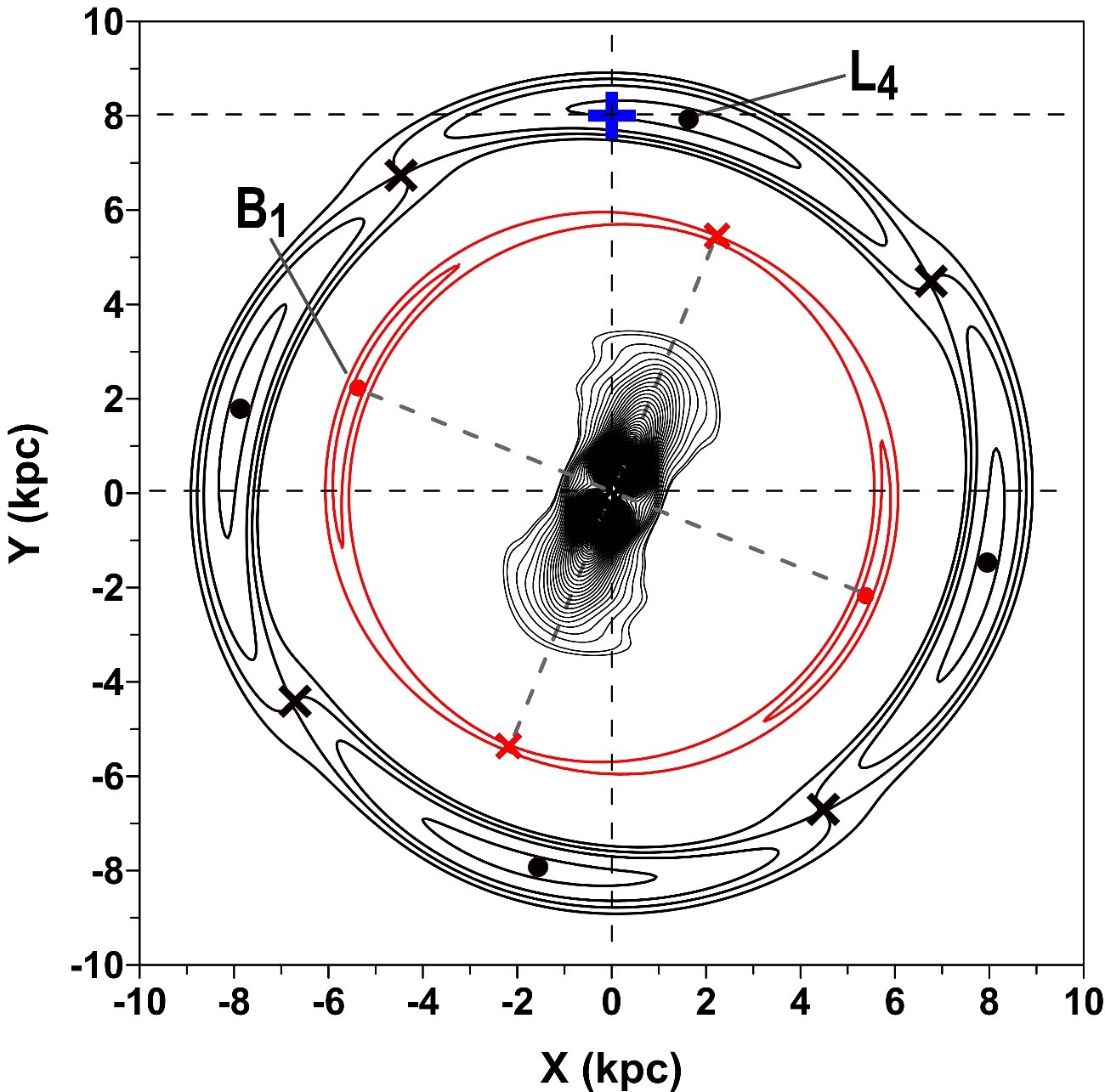,width=0.99\columnwidth ,angle=0}
\caption{Corotation zones originated by the four-arm spiral perturbation, with $\Omega_{\rm sp}=28.5$\,km\,s$^{-1}$\,kpc$^{-1}$ (black), and by the central bar perturbation, with $\Omega_{\rm bar}=40.0$\,km\,s$^{-1}$\,kpc$^{-1}$ (red). The different pattern speeds were chosen with the purpose of showing each contribution separately. The values of other parameters were taken from Table\,\ref{tab:1}. The zone of influence of the bar is centred at the origin; the directions of the major and minor axes of the bar are shown by dashed lines. The positions of the stable corotation centres are shown by dots, while the positions of the unstable saddle points are shown by crosses (see text for details). The position of the Sun is shown by a blue cross symbol.}
\label{fig:corotation}
\end{center}
\end{figure}

The topological portrait of the perturbation potential ${\mathcal H}_1(R,\varphi)$ in Eq.\,(\ref{eq:eq1-p}) is illustrated by its levels on the $X$--$Y$ plane in Fig.\,\ref{fig:U1}.
We can distinguish three topologically different regions: the inner region where the perturbation due to the bar is dominating, the intermediate region where the spiral perturbation  is beginning to gather strength, but is still strongly perturbed by the bar, and, finally, the outer region where the spiral perturbation is largely dominating.

In order to outline the inner region, which we refer to as the \emph{zone of influence of the bar}, we plot the energy levels in thick black lines in Fig.\,\ref{fig:U1}. The zone of influence extends up to $\sim$3.5\,kpc along the main axis of the bar, that is, beyond the physical extension of the bar with $R_{\rm bar}=2.9$\,kpc (see Table\,\ref{tab:1}).  The geometry of the zone of influence also differs from that of an elongated ellipsoid which shapes the bar (see Fig.\,\ref{fig:sketch}); it is rather like an observable box-shaped feature.

The other two regions are shown by the levels in grey lines in Fig.\,\ref{fig:U1}. In the intermediate zone, the loci of the azimuthal minima of the potential ${\mathcal H}_1(R,\varphi)$ (blue curves), which, outside the zone of influence of the bar, correspond to the loci of the main spiral arms in our model, are strongly perturbed; this is noted from the comparison with the unperturbed spiral arms shown in Fig.\,\ref{fig:sketch}. The two arms connected with the extremes of the bar, where the bar's force function is strongest, suffer significant  deformation, while the other two arms are partially vanishing. We estimate that the intermediate zone extends up to $\sim$4.3\,kpc until its effects disappear. It is interesting to note that the upper boundary of the intermediate zone closely matches the position of the ``extended bar'', which some authors observe at Galactocentric distances of 4--4.5 kpc \citep[e.g.][]{Lopez_CorredoiraEtal2007}.

Far away from the central bar/bulge structure, in the outer region of the Galactic disc, the amplitude of the bar/bulge perturbation decays and is several times smaller when compared to the strength of the spirals. It is expected that the corotation zone located close to the Sun will be only slightly affected by the bar/bulge perturbation. We note that this conclusion is only valid for the model calculated with parameters shown in Table\,\ref{tab:1} and will be tested with different values of the physical parameters of the bar/bulge in the following section.

Figure\,\ref{fig:corotation} shows the corotation zones originated by the gravitational perturbations due to the spiral arms and the bar/bulge structure. The corotation zones are visualised separately by plotting the levels of the effective potential (e.g. Paper I),
\begin{equation}\label{eq:effective}
\Phi_{\rm eff}(R,\varphi)= \Phi_0(R) + {\mathcal H}_1(R,\varphi)-\frac{1}{2}\Omega_p^2R^2,
\end{equation}
for each component of the term ${\mathcal H}_1(R,\varphi)$. It is worth emphasising that, for a given rotation curve (Eq. \ref{eq:Vrot}), the location of the corotation zone depends mostly on the pattern speed, and only slightly on details of the perturbing potential. In order to show each contribution separately, different pattern speeds have been chosen.

The corotation zone of the spiral arm structure, rotating with the pattern speed of $28.5$\,km\,s$^{-1}$\,kpc$^{-1}$, is shown by black levels in Fig.\,\ref{fig:corotation}. We can see four banana-like islands produced by four spiral arms; the black dots show the stable centres of the corotation zones characterised by the maxima of the effective potential (\ref{eq:effective}) in the absence of the bar, at $R=8.07$\,kpc and $\varphi=78^\circ .5$. The centre of the local corotation zone, in which the Sun is located (blue cross)is referred to hereafter as an $L_4$--centre. It should be emphasised that, throughout this paper, the position of the $L_4$--centre is calculated in the presence of the bar structure, unless stated otherwise. The stable corotation zones are separated by the unstable saddle points, shown by black crosses; these points are generally surrounded by orbits of unstable or chaotic motion.

The zone of influence of the bar shown by the levels of $\mathcal{H}_1(R,\varphi)$ is centred at the origin in Fig.\,\ref{fig:corotation}. Avoiding superimposing the contributions of the bar and of the arms, we assume the bar's pattern speed of $40.0$\,km\,s$^{-1}$\,kpc$^{-1}$, and the corotation radius is then obtained as $R=5.81$\,kpc. The bar perturbation generates two stable corotation centres (red points), located on the minor axis at opposite sides of the zone of influence of the bar. Each centre is surrounded by the banana-like domain of stable motion (red levels), which becomes unstable in the proximity of the two unstable saddle points located on the major axis of the bar (red crosses). We note that one unstable corotation point lies in almost the same direction as the $L_4$--centre, with a lag of $\sim 11^\circ .3$.  Hereafter, we refer to one of the stable centres generated by the bar as a $B_1$--centre.

It is worth emphasising that, in order to construct Fig.\,\ref{fig:corotation}, the corotation of each component of the perturbation was calculated independently from the others. In the case when all components are interacting between themselves, it is necessary to resolve Eqs.\,(\ref{eq:eq1})--(\ref{eq:eq1-p}) in order to decipher which one of the two stable points, $L_4$-- or $B_1$--centres, is dominating, particularly, in the solar neighbourhood.

\begin{figure}
\begin{center}
\epsfig{figure=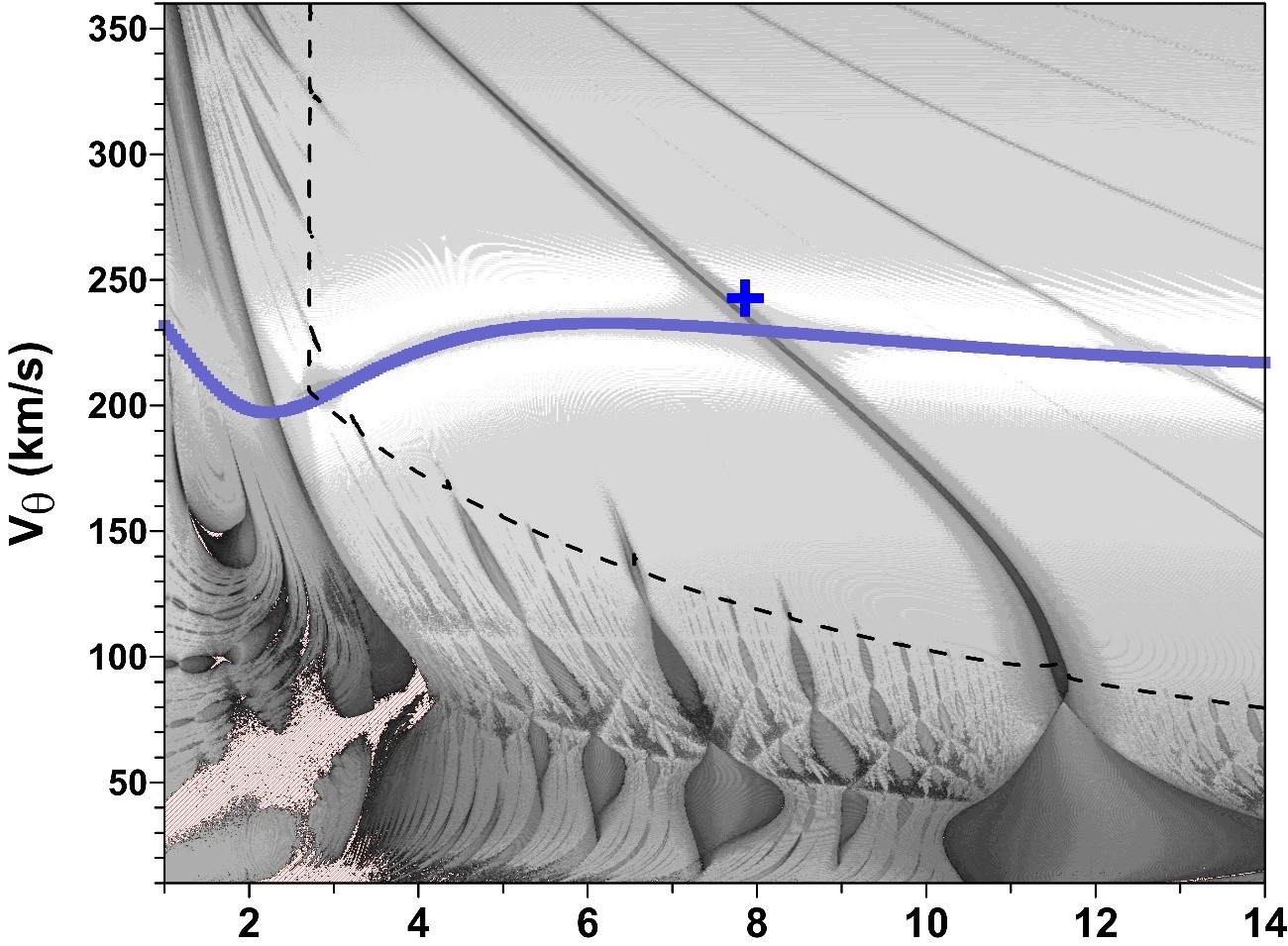,width=0.99\columnwidth ,angle=0}
\epsfig{figure=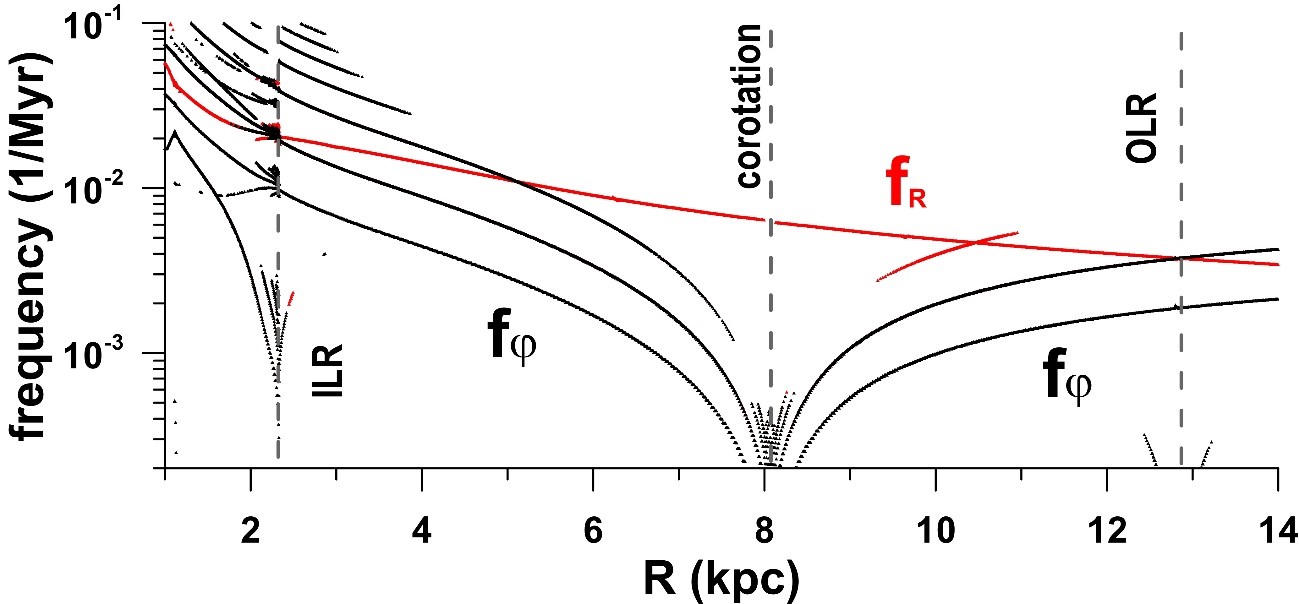,width=0.99\columnwidth ,angle=0}
\caption{Top: Dynamical map on the $R$--$V_\theta$ plane constructed with only the bar/bulge perturbation, for the basic set of parameters (Table\,\ref{tab:1}) and the initial $p_R=0$ and $\varphi=78^\circ .8$ (see text). The light grey tones represent regular orbits,
while increasingly dark tones correspond to increasing instabilities and chaotic motion. The  red-hatched regions contain strongly unstable and escaping orbits. The rotation curve is shown by a blue line. A dashed line delimits the initial conditions whose trajectories are inside or cross the central bar. The position of the Sun is shown by a blue cross symbol.
Bottom: Dynamical spectrum calculated with the bar model, along with a rotation curve shown by the continuous blue curve in the top panel. The proper frequency of the azimuthal oscillation, $f_\varphi$, and its harmonics are represented by black lines, while the frequency of the radial oscillation, $f_R$, is represented by a red line. The location of the corotation, ILR and OLR, is indicated by vertical dashed lines.
}
\label{fig:map-bar-spectr}
\end{center}
\end{figure}


\section{Dynamical maps on the $R$--$V_\theta$ plane }\label{sec:dyn-map}


In this section, the  main dynamical  features  of stellar motion in the equatorial plane of the Galaxy modelled by Eq.\,(\ref{eq:eq1}) are visualised on the representative $R$--$V_\theta$ plane  of  initial conditions, where the stellar azimuthal velocity (measured with respect to the inertial frame) is defined as \mbox{$V_\theta=L_z/R$}. To construct this plane, we fix the initial values of the momentum $p_R$ at zero. Indeed, all bounded orbits must have at least two turning points, defined by the condition $p_R=0$. We can also fix the initial value of the azimuthal angle $\varphi$. Indeed, we know that this angle is generally circulating; it oscillates only when the system is inside a corotation zone, particularly, inside the $L_4$--corotation shown in Fig.\,\ref{fig:corotation}. For the basic set of parameters (Table\,\ref{tab:1}), the position of the $L_4$--centre given by the effective potential (\ref{eq:effective}) is $R=8.08$\,kpc and $\varphi=78^\circ .8$; therefore, without loss of generality, the angular variable $\varphi$ can be initially fixed at $78^\circ 8$.

The first map presents dynamical features produced by the central bar/bulge structure alone (the pattern speed in this case is the rotation speed of the bar), with masses of $10^9$/$10^{10}$ solar masses (Fig. 6). The top graph shows the dynamical map and the rotation curve (blue line), while the bottom graph shows the dynamical spectrum calculated along the rotation curve (details on the construction of dynamical maps and spectra can be found in Appendix B). The interpretation of the map is simple: lighter grey tones represent regular quasi-periodic orbits, while increasingly dark tones correspond to increasing instabilities and chaotic motion. The resonances are then recognised as dark structures on the map, since the chaotic motion is associated to resonance separatrices. Finally, periodic orbits appear as white strips on the dynamical map.

The dynamical spectrum allows us to identify the nature of resonances while analysing the behaviour of the proper frequencies, $f_R$ and $f_\varphi$, which are frequencies of the radial and azimuthal oscillations, respectively. When $f_\varphi$ tends to zero, we have the corotation resonance; when $f_R-n\,f_\varphi\cong 0$ ($n$ is a simple integer), we have  one of the Lindblad resonances (see detailed description in Paper I). That is exactly what we observe in Fig.\,\ref{fig:map-bar-spectr}\,bottom: $f_\varphi$ (and its harmonics, black curves) tends to zero at $R=8.06$\,kpc, indicating the corotation zone, while a beating between $f_R$ (red curve) and $2\,f_\varphi$ occurs at $R=2.13$\,kpc (inside the zone of influence of the bar) and at $R=12.9$\,kpc, creating the ILR and OLR, respectively. There are also two beatings between third harmonic $3\,f_\varphi$ and $f_R$ (3/1 resonance), but their dynamical effects are too weak to be observable on the spectrum and on the map.

According to our model given by the Hamiltonian (\ref{eq:eq1}), in the absence of the spiral perturbation, the exact position of the stable corotation $B_1$--centre of the bar/bulge, rotating with the pattern speed $\Omega_p=28.5$\,km\,s$^{-1}$\,kpc$^{-1}$, is $R=8.06$\,kpc and $\varphi=157^\circ.5$. It is far enough from the Sun to be visible on the map constructed with the fixed $\varphi=78^\circ .8$. On the other hand, the unstable saddle point of the bar's corotation is located at $\varphi=67^\circ.5$ (see Fig.\,\ref{fig:corotation}), in the neighbourhood of the Sun. As a consequence, we observe instabilities associated to this point on the map at the top of Fig.\,\ref{fig:map-bar-spectr}; these instabilities are extended continuously beyond the Sun's neighbourhood, in the form of a thick strip, for all  $V_\theta$--values. The OLR also appears on the map in a similar way, as a very thin strip, at large Galactic radii.

On the other hand, the ILR has dominating effects in the inner region of the disc. The initial conditions of the resonant orbits lie inside the zone influence of the bar, which is delimited by a dashed line in Fig.\,\ref{fig:map-bar-spectr}\,(top); therefore, the detailed analysis of the resonant dynamics will require a more realistic model for the bar/bulge structure.  However, some generic dynamical features inside this zone should be emphasised in advance: First, all stellar orbits starting with low azimuthal velocities, $V_\theta < 100$\,km\,s$^{-1}$, are orbits which either evolve inside or cross the zone of influence of the bar, independently of the initial radial distance. 
Second, for the mass of the bar/bulge of $10^9$/$10^{10}$ solar masses, the near-circular orbits starting inside the bar and close to the rotation curve (blue continuous curve) are very regular, as is the  stellar motion of the objects with higher azimuthal velocities. Finally, the bar's perturbation enhances the resonant effects on these `interior' orbits, even in the case of high-order resonances, forming a complex resonance web seen in the region below the dashed curve in the top part of Fig.\,\ref{fig:map-bar-spectr}. According to the theories of resonant motion, depending on initial configurations, some of these resonances could protect the stellar motion from collisions and escapes, or, on the contrary, could provoke strong dynamical instabilities. One small domain of such instabilities is  shown by a red-hatched region in this latter figure.
\begin{figure}
\begin{center}
\epsfig{figure=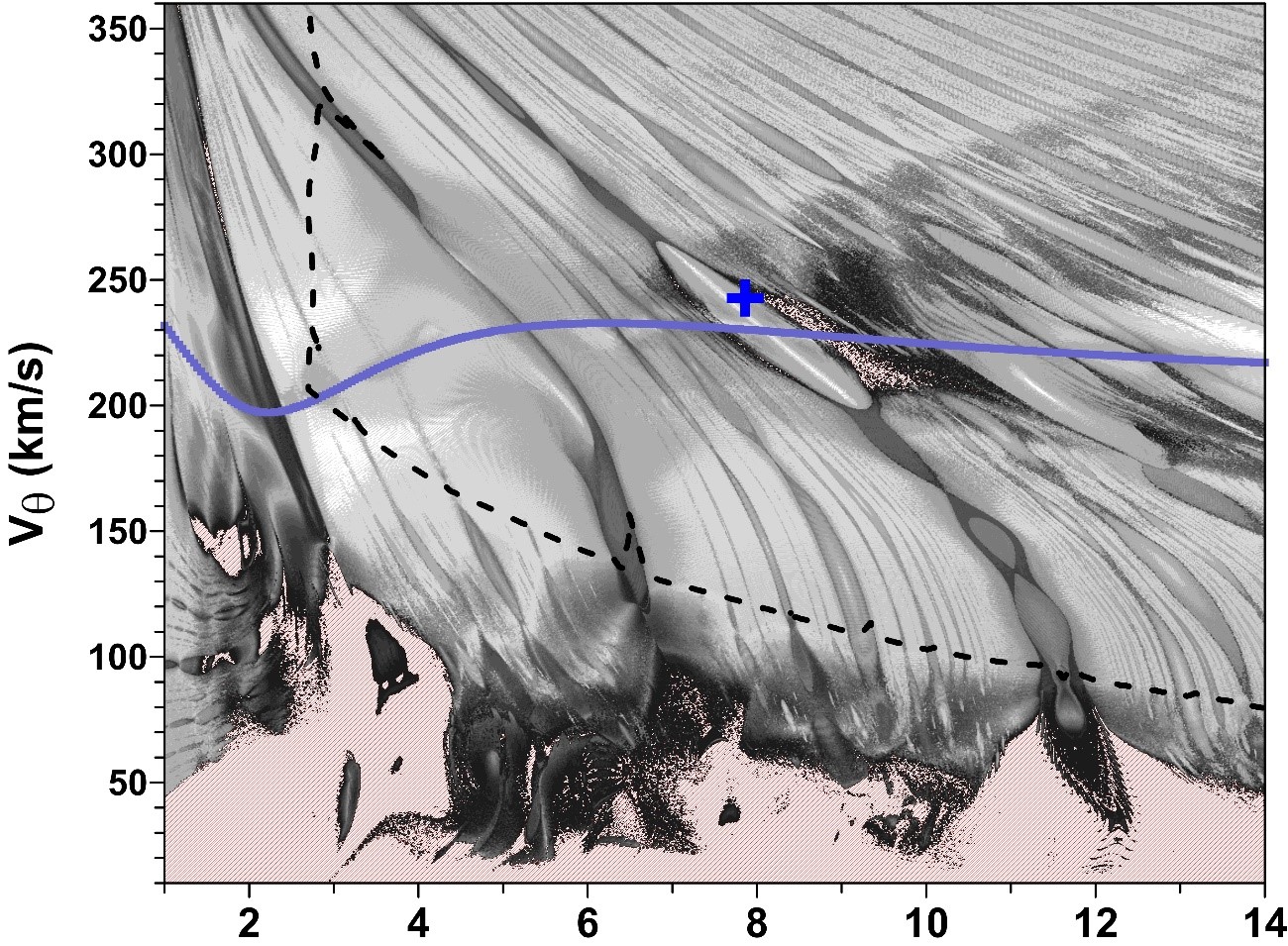,width=0.99\columnwidth ,angle=0}
\epsfig{figure=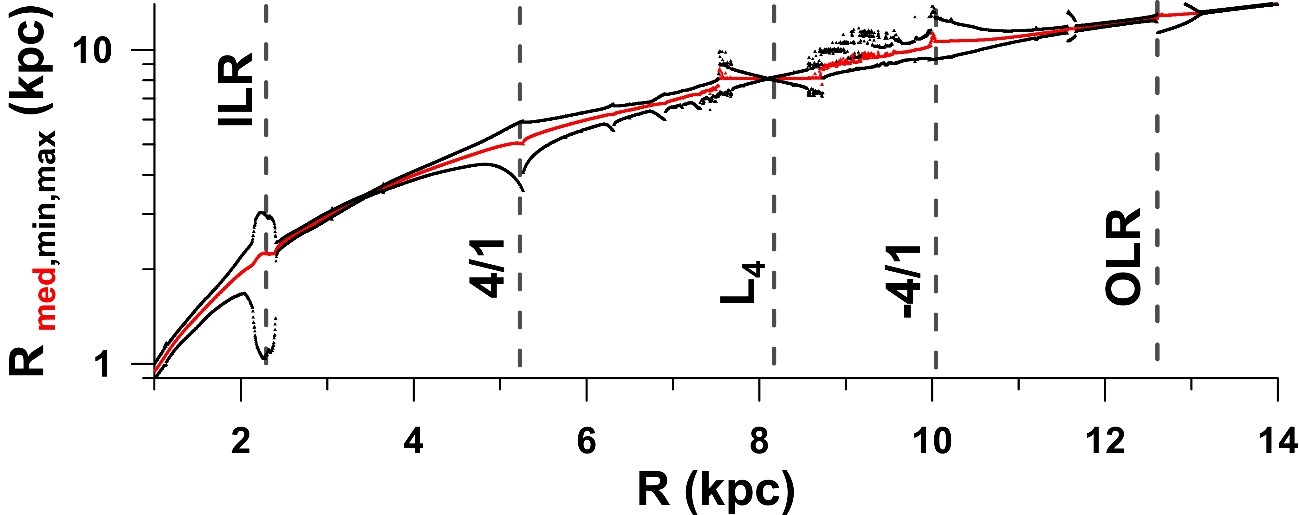,width=0.99\columnwidth ,angle=0}
\caption{Top: Same as in Figure\,\ref{fig:map-bar-spectr}\,top, except adding the spiral arms perturbation.
Bottom: The averaged (red) and maximal/minimal (black) values of the $R$--variable calculated over 10\,Gyr as function of initial values of $R$. The locations of the $L_4$--corotation and the  strongest resonances are indicated by vertical dashed lines.
}
\label{fig:map-all-mass09}
\end{center}
\end{figure}

The second map presented in this section in Fig.\,\ref{fig:map-all-mass09} is constructed with the same initial conditions as the first map, except that the spiral arms' perturbation is added in the model, with parameters taken from Table\,\ref{tab:1}. The spiral perturbation on the stellar motion, rotating with the same pattern speed, enhances the main dynamical features already seen on the map in Fig.\,\ref{fig:map-bar-spectr}. The resonances outside the zone of influence of the bar gather strength (e.g. the 4/1 ILR which intersects the rotation curve at $R=5.2$\,kpc), while the low-velocity motions below the dashed curve amplify their instabilities. There is only one qualitative difference between the two maps: the appearance of the stable corotation region in the Sun's neighbourhood associated to the spiral arms.

Using the model given by the Hamiltonian (\ref{eq:eq1}), we calculate the global equilibrium for the basic set of parameters from Table\,\ref{tab:1} and obtain its exact position at $R=8.08$\,kpc and $\varphi=78^\circ.8$. This is the $L_4$--centre of the local corotation zone, and both the local arm and the Sun are situated inside this region (see Paper II). The presence of the stable corotation zone surrounded by the thick layer of instability is the central feature of the map in Fig.\,\ref{fig:map-all-mass09}.

The behaviour of the orbits starting with velocities along the rotation curve (continuous blue curve) can be observed in the bottom part of Fig.\,\ref{fig:map-all-mass09}, where we plot the averaged values (red) and maximal/minimal variations (black) of the radial distance of the stars, obtained over 10\,Gyr, as a function of the initial values of $R$. As expected, near-circular orbits oscillate with very small amplitude, unless the motion occurs inside a resonance, where oscillation amplitude is amplified. This property allows us to observe passages of the initial conditions through several resonances  in Fig.\,\ref{fig:map-all-mass09}. We note that the vertical axis in the graph is in logarithmic scale, in which the excitation due to the ILR (at $R=2.13$\,kpc) seems to be most prominent. Even in this exceptional case, the trajectories of near-circular objects starting inside the bar are confined to inside the bar, varying between 1 and 3\,kpc.

\begin{figure}
\begin{center}
\epsfig{figure=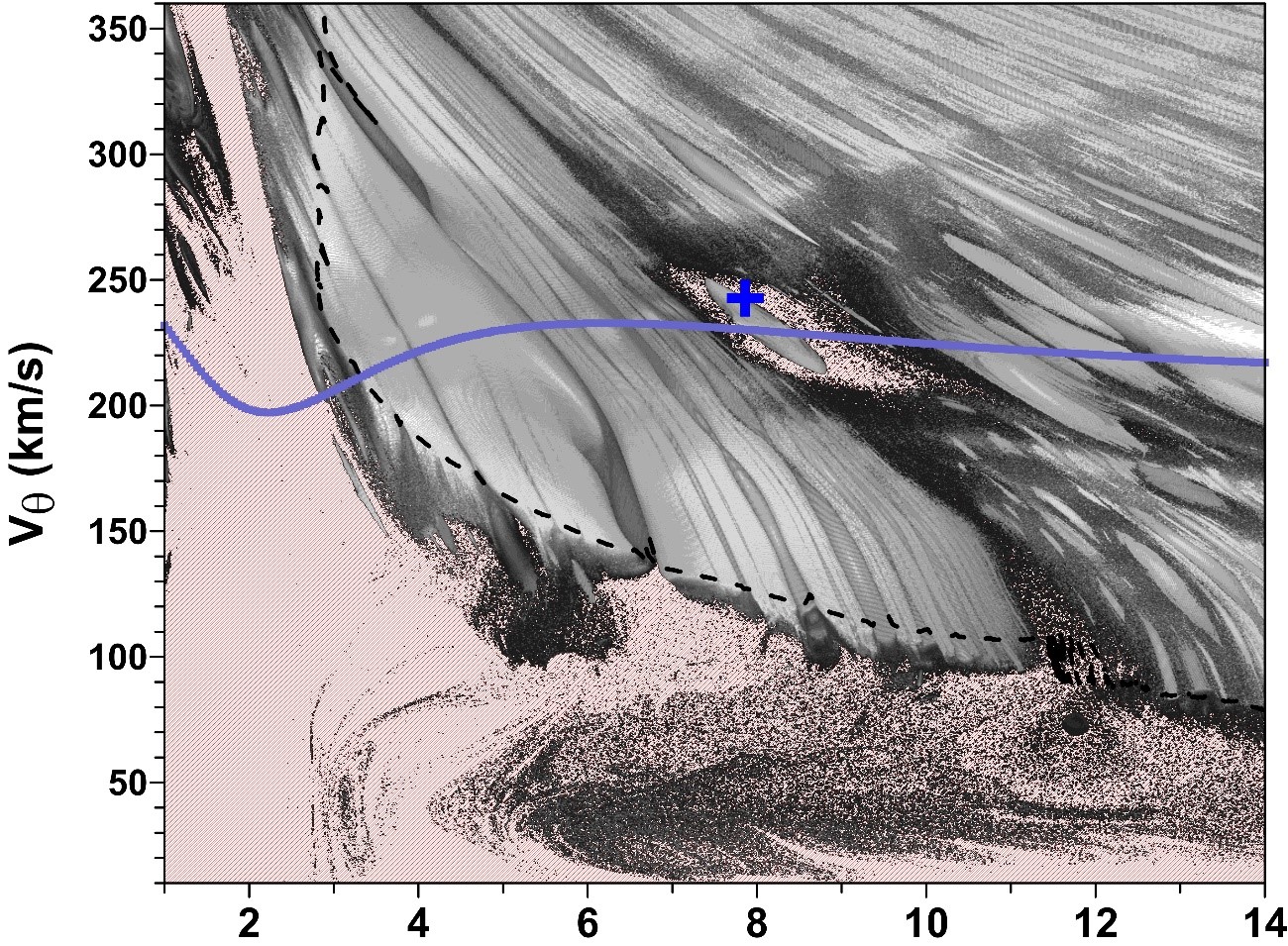,width=0.99\columnwidth ,angle=0}
\epsfig{figure=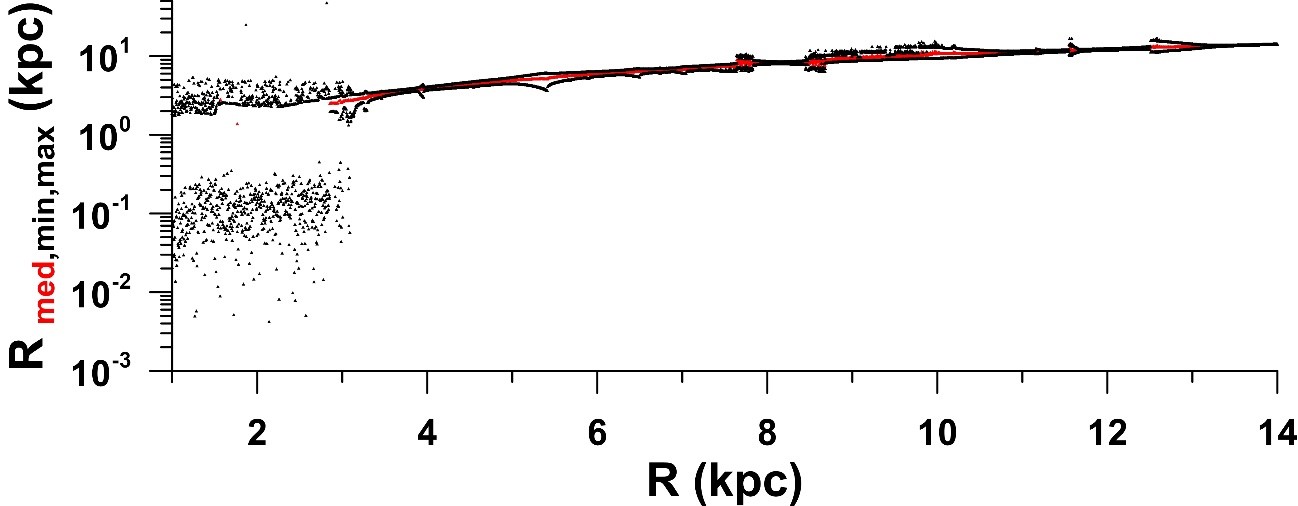,width=0.99\columnwidth ,angle=0}
\caption{Same as in Figure\,\ref{fig:map-all-mass09}, except, for the bar's mass equal to $1\times 10^{10}$ solar masses.
}
\label{fig:map-all-mass10}
\end{center}
\end{figure}


\subsection{Dynamical constraints on the bar's mass}\label{sec:map-mass}

All results shown in the previous sections were obtained with the total mass of the central bar fixed at $10^9$ solar masses. In this section, we look for constraints on the bar's mass analysing its dynamical effects on the stellar motion in the spiral galaxy described by the parameters from Table\,\ref{tab:1}. Once again, we reiterate that the mass of the bulge cannot be constrained by analysing the dynamics in the equatorial plane.
The first step is to recalculate the dynamical map in Fig.\,\ref{fig:map-all-mass09}, using the same initial conditions and parameters, except for the bar's mass, which is increased tenfold. The top part of Fig.\,\ref{fig:map-all-mass10} shows the dynamical map that is obtained. The comparison between the two maps reveals that perturbation due to the growing bar's mass affects the neighbourhood of the Sun, increasing the layers of instabilities. This effect is a consequence of the fact that the unstable centre of the bar's corotation, rotating with the same speed of the spiral arms, is only  $11^\circ .3$ lagged from the stable $L_4$--centre. The Sun is located now outside the corotation zone and its motion is chaotic.
\begin{figure}
\begin{center}
\epsfig{figure=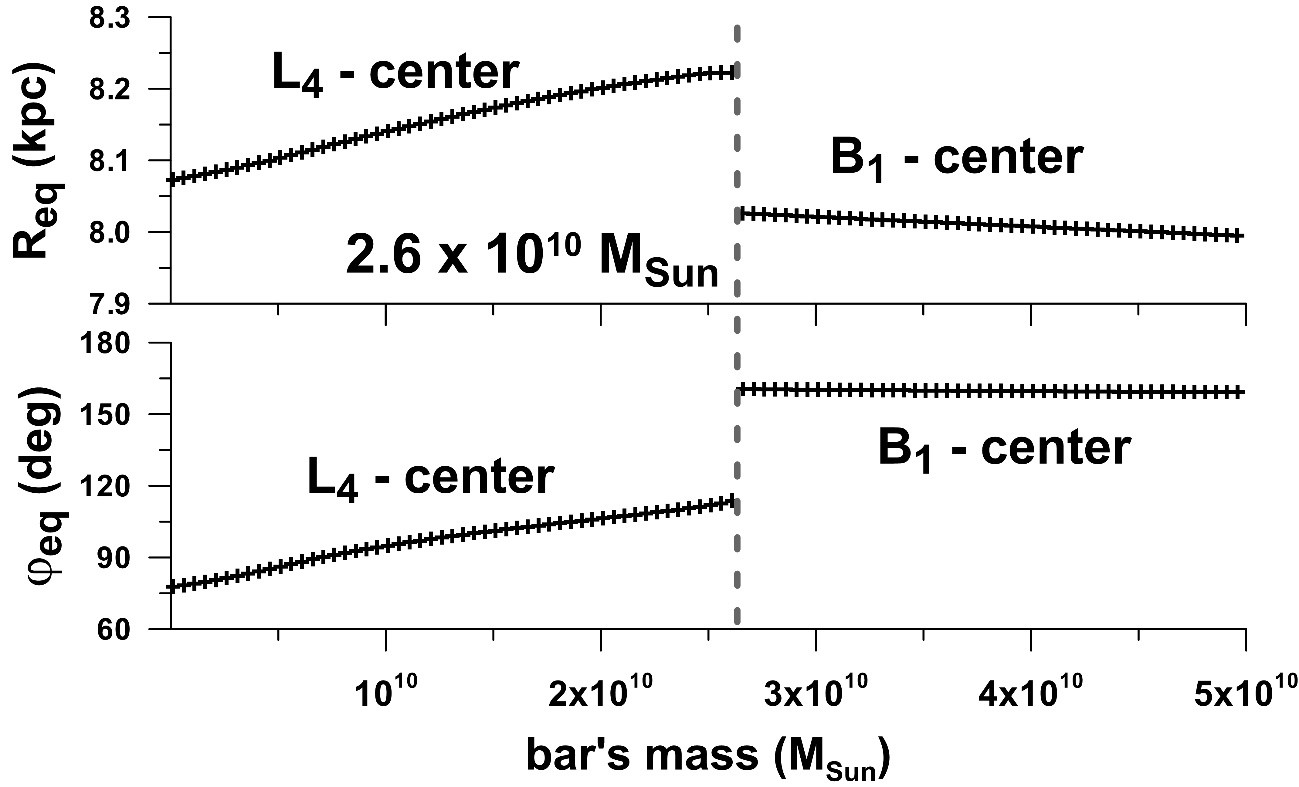,width=0.99\columnwidth ,angle=0}
\caption{The evolution of the radius (top) and phase (bottom) of the global maximum of the effective potential (\ref{eq:effective}) as a function of the mass of the central bar. For smaller values of the bar's mass, the equilibrium solution is associated to the stable $L_4$--centre of corotation. Its monotonous evolution with the increasing mass is interrupted at $M_{\rm bar}=2.6\times 10^{10}M_\odot$,  when $L_4$--centre becomes unstable and $B_1$--centre has maximal energy (see Fig.\,\ref{fig:corotation}).
}
\label{fig:extrema-GMbar}
\end{center}
\end{figure}

In the following, we analyse the evolution of the $L_4$--centre with the growing bar's mass. We obtain the location of the global maximum of the effective potential (\ref{eq:effective}) in the equatorial Galactic plane, for different values of the mass of the bar, but keeping the parameters from Table\,\ref{tab:1}. The solutions obtained are shown in Fig.\,\ref{fig:extrema-GMbar}, where the top graph shows the radial distance $R_{\rm eq}$ and the bottom graph shows the azimuthal angle $\varphi_{\rm eq}$ of the global maximum as functions of the bar's mass from the range between $1\times 10^{8}M_\odot$ and $5\times 10^{10}M_\odot$. When the bar's mass is small, the spiral perturbations are dominating and, consequently, the $L_4$--centre of the spiral corotation is a maximum of the effective potential (\ref{eq:effective}). When the bar's mass is increasing continuously, the $L_4$--centre is dislocated slightly from its initial position at $R=8.06$\,kpc and $\varphi=76^\circ$  (defined in the absence of the bar). This shows that the position of the $L_4$--centre of corotation depends only slightly on the bar's parameters.

Figure\,\ref{fig:extrema-GMbar} shows that the continuous evolution of the $L_4$--centre with the increasing mass of the bar is suddenly interrupted when $M_{\rm bar}$ reaches $\sim 2.6\times 10^{10}M_\odot$; the perturbation of the bar at this instant becomes dominating and the $B_1$--centre of the bar's corotation assumes the role of the global maximum of the effective potential (\ref{eq:effective}). When $M_{\rm bar}$ continues increasing, the corotation radius of the $B_1$--centre decreases very slightly, while its phase, aligned with the minor axis of the bar's ellipsoid, remains the same.
\begin{figure}
\begin{center}
\epsfig{figure=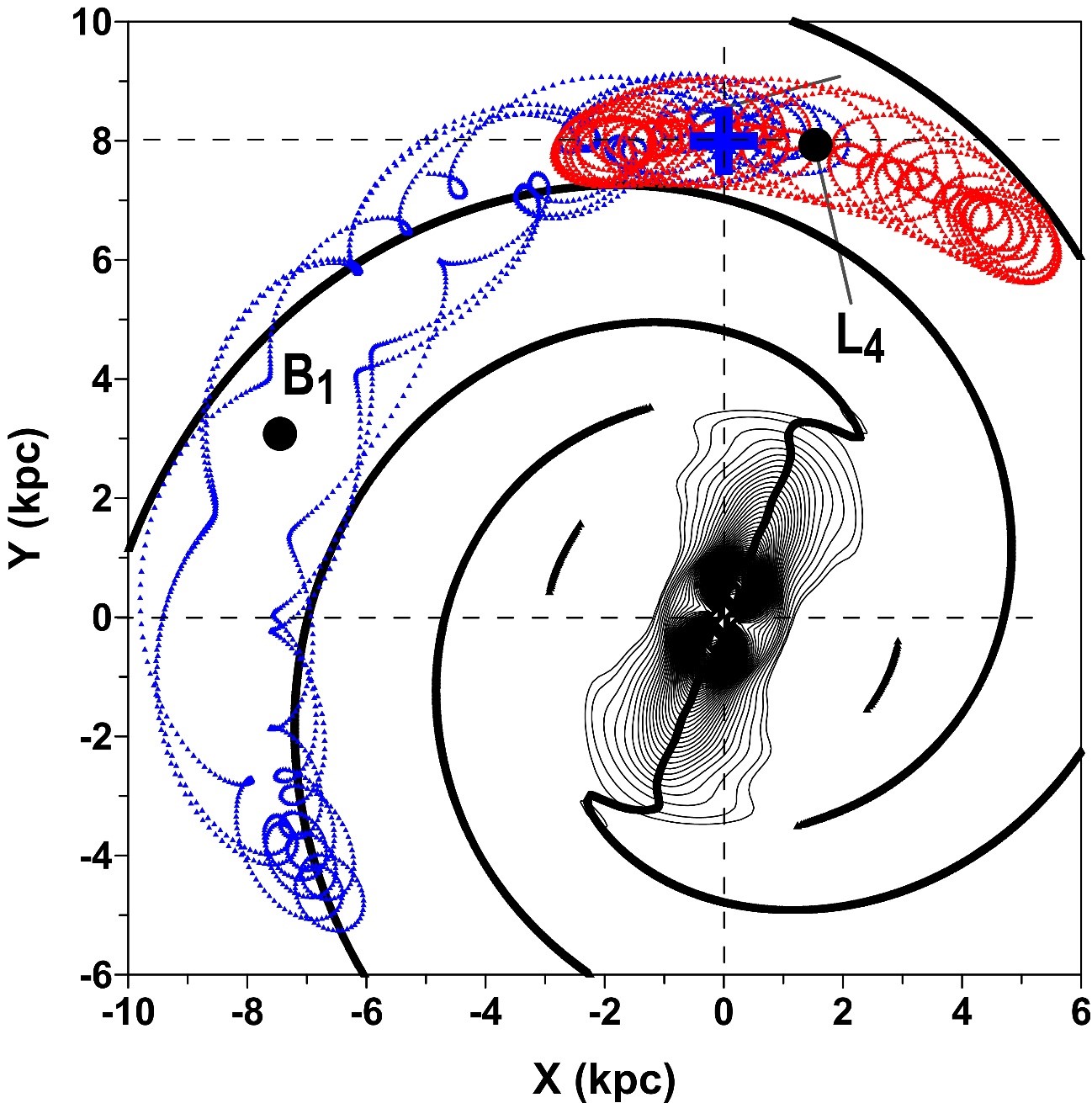,width=0.8\columnwidth ,angle=0}
\caption{Two projections of the Sun's orbit on the $X$-$Y$ plane, calculated with $M_{\rm bar}=1\times 10^{9}M_\odot$ (red) and $M_{\rm bar}=3\times 10^{10}M_\odot$ (blue), over 5\,Gyr; all other parameters are taken from Table\,\ref{tab:1}. The initial position of the Sun is shown by a blue cross. For the smaller bar's mass, the Sun's trajectory oscillates around the $L_4$--centre, while, for the larger mass, it oscillates around  the $B_1$--centre. The zone of influence of the bar/bulge and the loci of the spiral arms are shown by black curves.
}
\label{fig:sun-orbits}
\end{center}
\end{figure}

To understand the stellar dynamics in this case, we integrate the orbit of the Sun using two different values of $M_{\rm bar}$, $1\times10^9$ (red) and $3\times10^{10}$ (blue) solar masses, keeping the other parameters from Table\,\ref{tab:1}; the trajectories obtained are shown in Fig.\,\ref{fig:sun-orbits} by red and blue points, respectively. Both orbits, starting at the same initial configuration (a blue cross symbol), are librating; however, the red path librates around the $L_4$--centre of the spiral corotation, while the blue path librates around $B_1$--centre of the bar's corotation. The first orbit oscillates between the Sagittarius-Carina and Perseus arms, never crossing them; this resonant behaviour is characteristic of the objects from the local arm (see Paper II). On the other hand, the Sun's orbit evolving in the bar's corotation has a large amplitude of oscillation, crosses the Sagittarius-Carina and Crux-Centaurus arms, and shows irregular behaviour. From the point of view of conditions that are favourable to life, the dynamics of the Sun in the former case would provide a possible upper limit on the bar's mass.

Moreover, when we return to analyse the dynamical map in Fig.\,\ref{fig:map-all-mass10}, constructed with $M_{\rm bar}=1\times10^{10}M_\odot$, we note that almost the whole domain of the stellar orbits that start inside or cross the bar (initial conditions below the dashed line) is strongly unstable (red hatched domains). The graph in the bottom panel of Fig.\,\ref{fig:map-all-mass10} shows the variation of the orbits starting with near-circular velocities: we plot the averaged values (red) and maximal/minimal variations (black) of the radial distance of stars as a function of the initial values of $R$. We note very large radial excursions of objects starting inside the bar: in a few hundred million years, they are ejected from the bar's zone. It is expected that this behaviour will jeopardise the integrity of the bar structure as a whole. The quantitative analysis of constraints on the mass of the bar is presented in Sects.\,\ref{sec:bar-size} and \ref{sec:bar-speed}.


\section{Dependence on the parameters of the bar}\label{sec:map-param}

In this section, we test different physical and geometrical parameters of the bar, such as the bar's mass, flattening, radius and  initial orientation. The parameters from the basic set (Table\,\ref{tab:1}) are changed one-by-one, inside the ranges shown in Table\,\ref{tab:1-1}, in order to establish their possible limits.

\begin{figure}
\begin{center}
\epsfig{figure=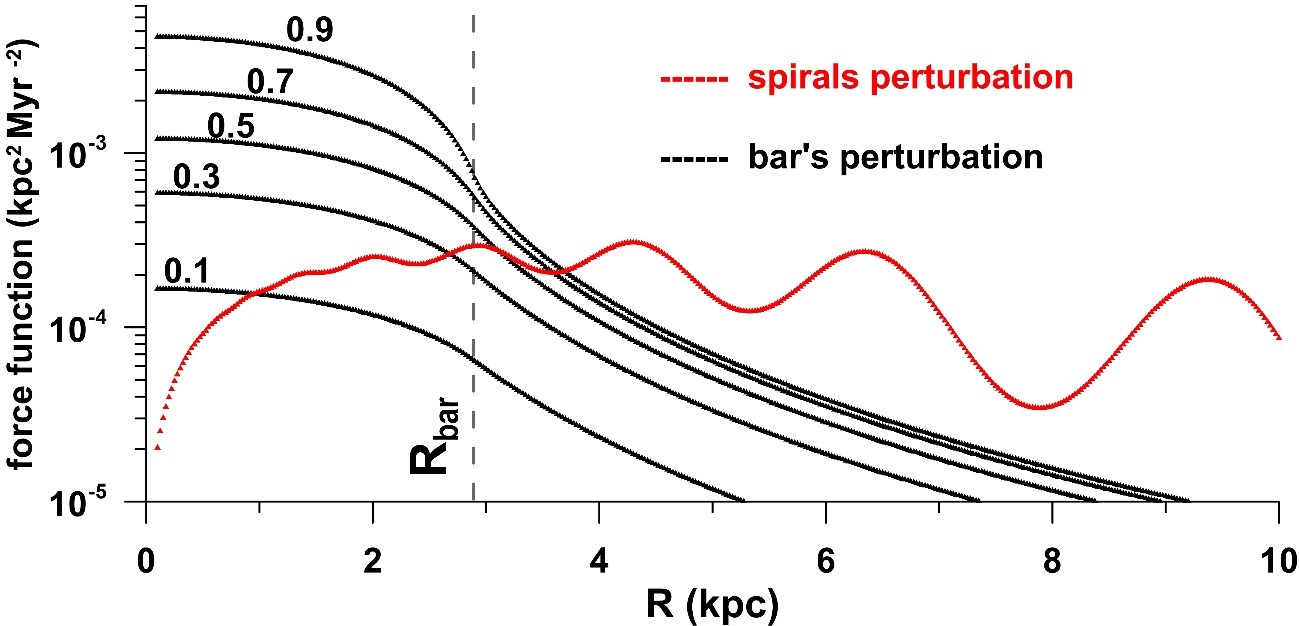,width=0.99\columnwidth ,angle=0}
\caption{Families of the force function of the bar (black) parameterized by the different values of the bar's flattening, from 0.1 to 0.9; the rest of the parameters is taken from Table\,\ref{tab:1}. The force function of the spiral arms is shown by a red curve.
}
\label{fig:var-flattening}
\end{center}
\end{figure}


\subsection{Bar's flattening}\label{sec:map-flat}

Analysing the bar's flattening is relatively simple. According to the definition in Eq.\,(\ref{eq:flat}), its possible values lie between 0 and 1.  We calculate the families of the force function of the bar  parameterized by the different values of $f_{\rm bar}$, from 0.1 to 0.9. All functions were calculated along the major axis of the bar fixed at $\varphi=67^\circ.5$ (see Sect.\,\ref{sec:model-bar}), where the bar's perturbation is strongest; the parameters were taken from Table\,\ref{tab:1}.

Figure\,\ref{fig:var-flattening} shows the families as functions of the Galactocentric distance by black curves; for the sake of comparison, we also plot the force function of the spiral arms by a red curve. We note that the non-axisymmetric perturbation of the bar becomes stronger with the increasing  flattening of the bar, $f_{\rm bar}$. The domain of the overlap with the spiral perturbation is also increasing, which means that the inner and intermediate zone of the influence of the bar (see Sect.\,\ref{sec:topology}) are expanding. Indeed, for $f_{\rm bar}>0.5$, the force function of the bar dominates over the spirals, even beyond its physical extension given by $R_{\rm bar}=2.9$\,kpc.

However, Fig.\,\ref{fig:var-flattening} shows that, for $M_{\rm bar}=10^{9}M_\odot$, the zone of influence of the bar never approaches the domain of the spiral corotation, where the Sun is evolving. Therefore, it is expected that, for the adopted bar's mass, the motion of the Sun and the location of the $L_4$--centre on the $X$--$Y$ plane is only slightly affected by increasing $f_{\rm bar}$.

\begin{figure}
\begin{center}
\epsfig{figure=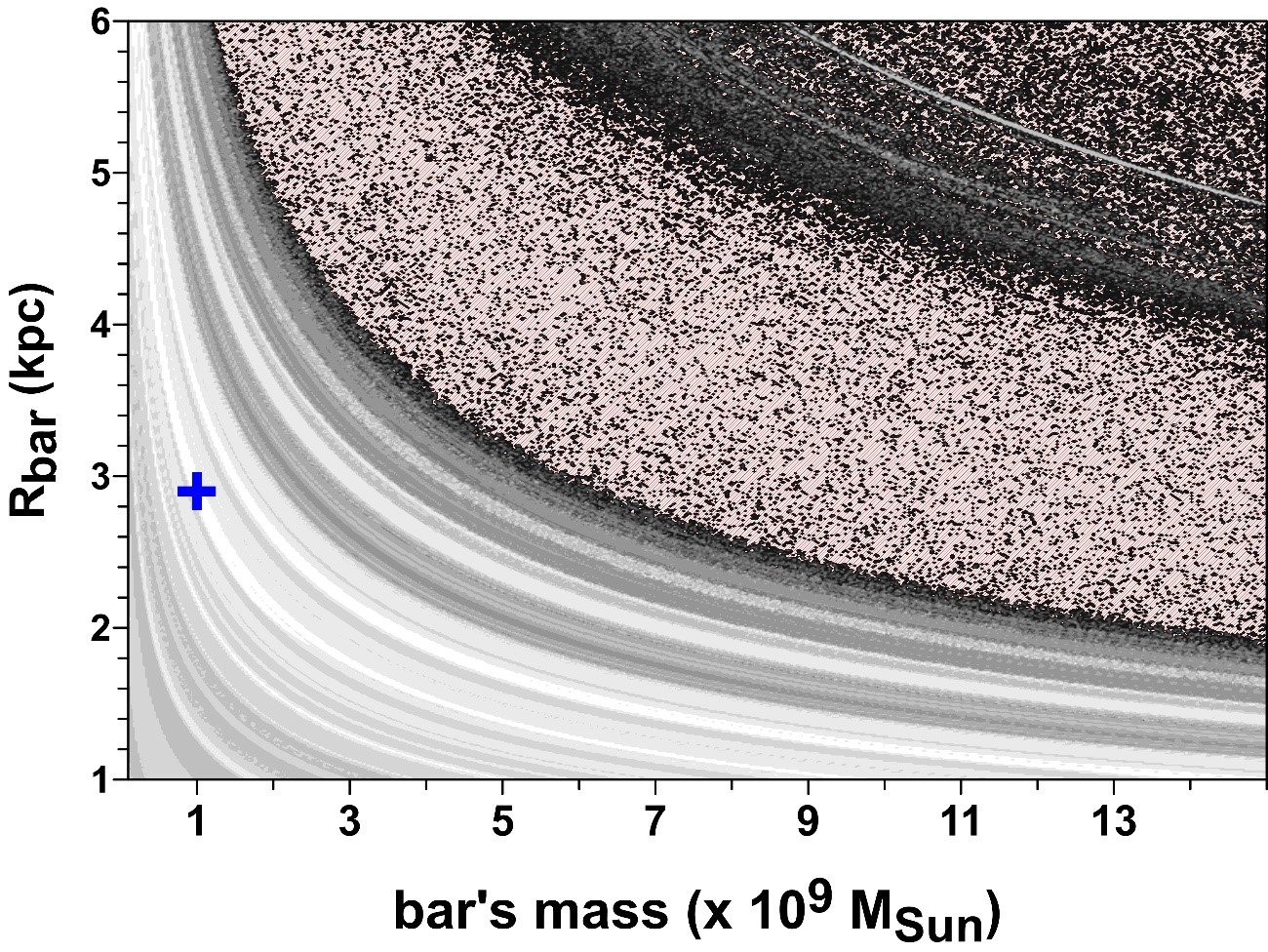,width=0.99\columnwidth ,angle=0}
\caption{Dynamical map on the parametric plane $M_{\rm bar}$--$R_{\rm bar}$ showing the stability of the Sun's orbits as a function of the bar's mass and radius. The orbits are stable in the light domains and are strongly unstable in the red hatched regions.  The parameters adopted in the basic set place the Sun in the position  shown by a blue cross symbol.
The values of the rest of the parameters are taken from Table\,\ref{tab:1}.
}
\label{fig:spectr-Rbar-Sun}
\end{center}
\end{figure}


\subsection{The bar's size and initial orientation}\label{sec:bar-size}

The results obtained in the previous sections are based on the adopted hypothesis that the initial orientation of the bar $\gamma^0_{\rm bar}$ defines its radius (see Fig.\,\ref{fig:barradius}). From observations, the bar's angle with respect to the Sun's direction varies in the range from 10$^\circ$  to 30$^\circ$ \citep{BobylevEtal2014}, which defines $\gamma^0_{\rm bar}$ in the range from 60$^\circ$ to 80$^\circ$ and, consequently, the bar's radius in the range from 2.78\,kpc to 3.05\,kpc. In this section, we relax this restriction and vary the bar's radius in the range from 1\,kpc to 6\,kpc, keeping $\gamma^0_{\rm bar}=67^\circ.5$.

We test the effects of the bar's size on the Sun's motion, which is representative of the dynamical stability of objects from the local arm. For this, we construct the dynamical map on the parametric plane $M_{\rm bar}$--$R_{\rm bar}$ shown in Fig.\,\ref{fig:spectr-Rbar-Sun}. Varying the bar's mass in the range between $10^{8}M_\odot$ and $1.5\times 10^{10}M_\odot$ and the bar's radius between 1\,kpc and 6\,kpc, we analyse the dynamical  stability of the Sun, with coordinates $X=0$ and $Y=8.0$\,kpc, and the velocities $p_R=-11$\,km\,s$^{-1}$ and $V_\theta=242.24$\,km\,s$^{-1}$. The rest of the parameters are taken from Table\,\ref{tab:1}.

The perturbations to stellar motion due to the bar are weak in the light-tone domain on the dynamical map in Fig.\,\ref{fig:spectr-Rbar-Sun}; they increase in the darker zones and provoke strong instabilities in the red-hatched region. (The fine effects in the solar motion shown by slight variations of grey tones in the domain of stable orbits are not analysed here.) The analysis of the map shows the effects of the mass of the bar and of its radius on the stability of the solar orbit: the increasing mass reduces the stability of the stellar motion inside the corotation zone, while the decreasing radius enhances this stability. Both parameters are saturated: the bar's mass at $\sim10^{9}M_\odot$ and the bar's radius at $\sim$1.7\,kpc. For the adopted value $R_{\rm bar}=2.9$\,kpc, the current motion of the Sun and objects from the local arm remains stable up to $M_{\rm bar}\approx 5\times 10^{9}M_\odot$.

The parameter $\gamma^0_{\rm bar}$ defines the initial orientation of the bar in the chosen reference frame and its value, fixed at $67^\circ.5$ (see Table\,\ref{tab:1}), is tightly constrained by observations \citep[e.g.][]{BobylevEtal2014}. In this configuration, the unstable saddle point of the bar's corotation approximately matches the position of the stable $L_4$--centre of the spiral corotation, if the rotation speed of the bar is equal to the speed of the spiral structure (see Fig.\,\ref{fig:corotation}). Indeed, in this case, the relative angular lag between these points is of only 11$^\circ$.3, and the corotation radii of the bar and the spiral arms are the same. It is clear that this situation could be unfavourable for the stability of the $L_4$--centre, since the instabilities caused by the bar' saddle point affect the $L_4$ corotation zone.
However, the calculations made throughout this paper show that, for $M_{\rm bar}< 5\times 10^{9}M_\odot$, the bar produces only insignificant modulations in the location of the corotation $L_4$--centre. What happens for the different values of the bar's speed is discussed in the following section. We see below that there is a correlation between the maximum value of $M_{\rm bar}$ and the maximum value of the bar's rotation speed $\Omega_{\rm bar}$: For masses of the order of a few $10^9 M_\odot$, we obtain upper limits on $\Omega_{\rm bar}$.

\begin{figure}
\begin{center}
\epsfig{figure=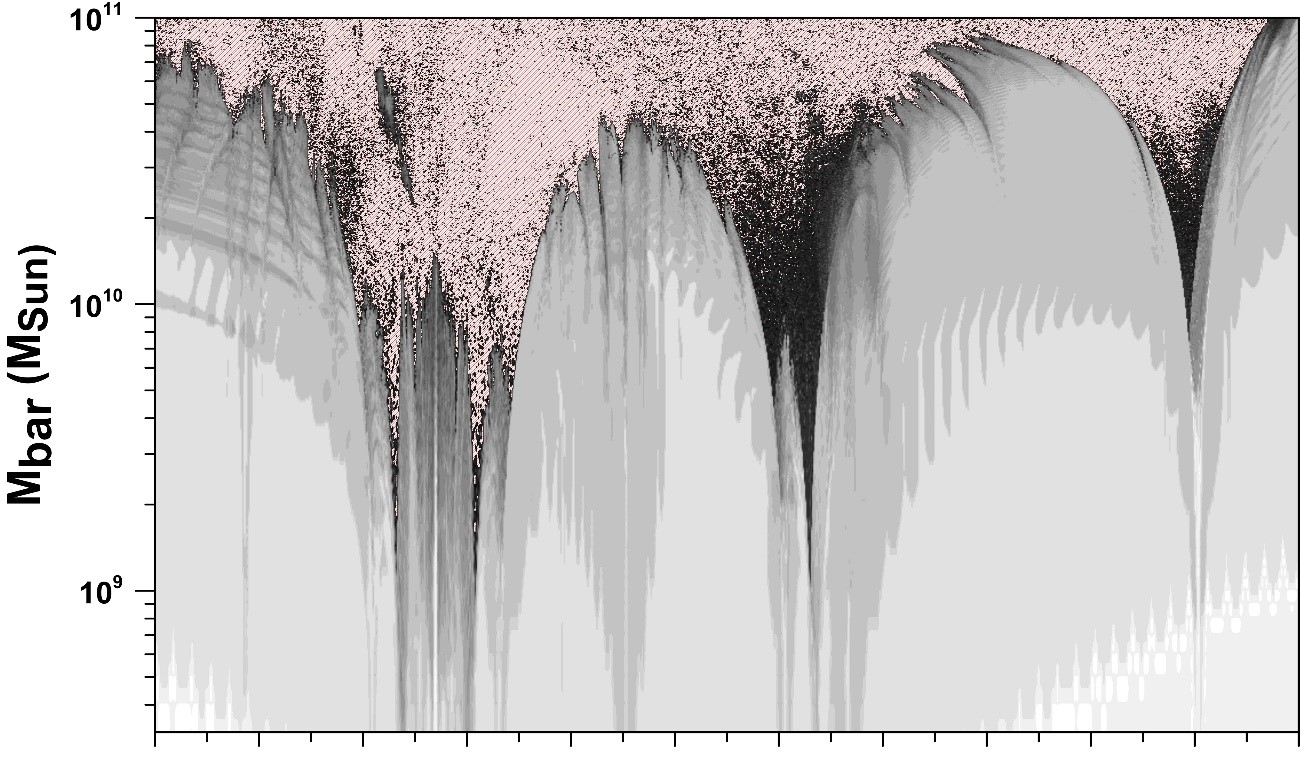,width=0.99\columnwidth ,angle=0}
\epsfig{figure=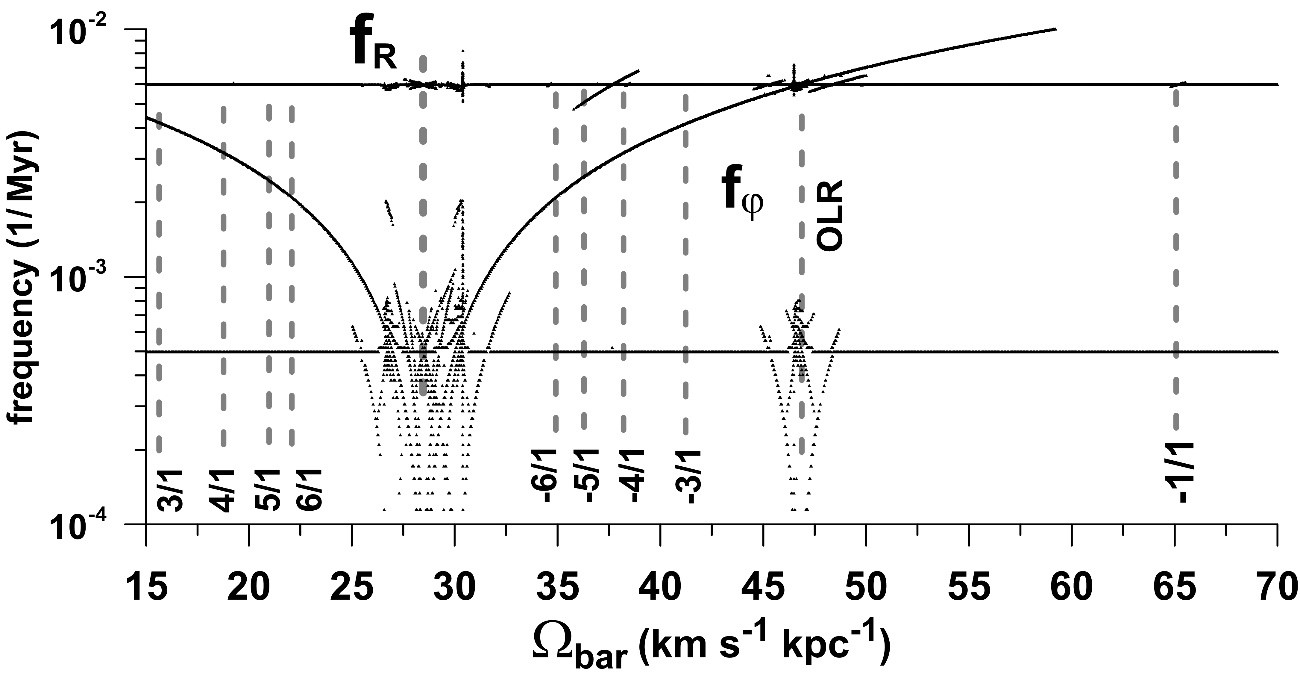,width=0.99\columnwidth ,angle=0}
\caption{\emph{Top:} Dynamical map on the parametric plane $\Omega_{\rm bar}$--$M_{\rm bar}$ for the $L_4$--center calculated with the parameters of Table\,\ref{tab:1} (representing the spiral corotation zone); $L_4$ is a fixed point only if $\Omega_{\rm bar} = \Omega_p$. We see a wide region of stability (light gray tones) for bar masses of the order of $10^9 M_\odot$ and for all values of $\Omega_{\rm bar}$, except in regions when the spiral corotation zone and the bar's main resonances overlap.  In this case, there appears a wide region of chaos (dark gray and black tones). For bar masses greater than $\sim 10^{10} M_\odot$, $L_4$ is strongly unstable for any value of $\Omega_{\rm bar}$ (red hatched region).
\emph{Bottom:} Dynamical spectrum of the $L_4$ orbit calculated for $M_{\rm bar} = 10^9 M_\odot$. The nominal values of the bar's Lindblad resonances are denoted by vertical dashed lines.
}
\label{fig:Omega-bar-GMbar-L4}
\end{center}
\end{figure}

\begin{figure}
\begin{center}
\epsfig{figure=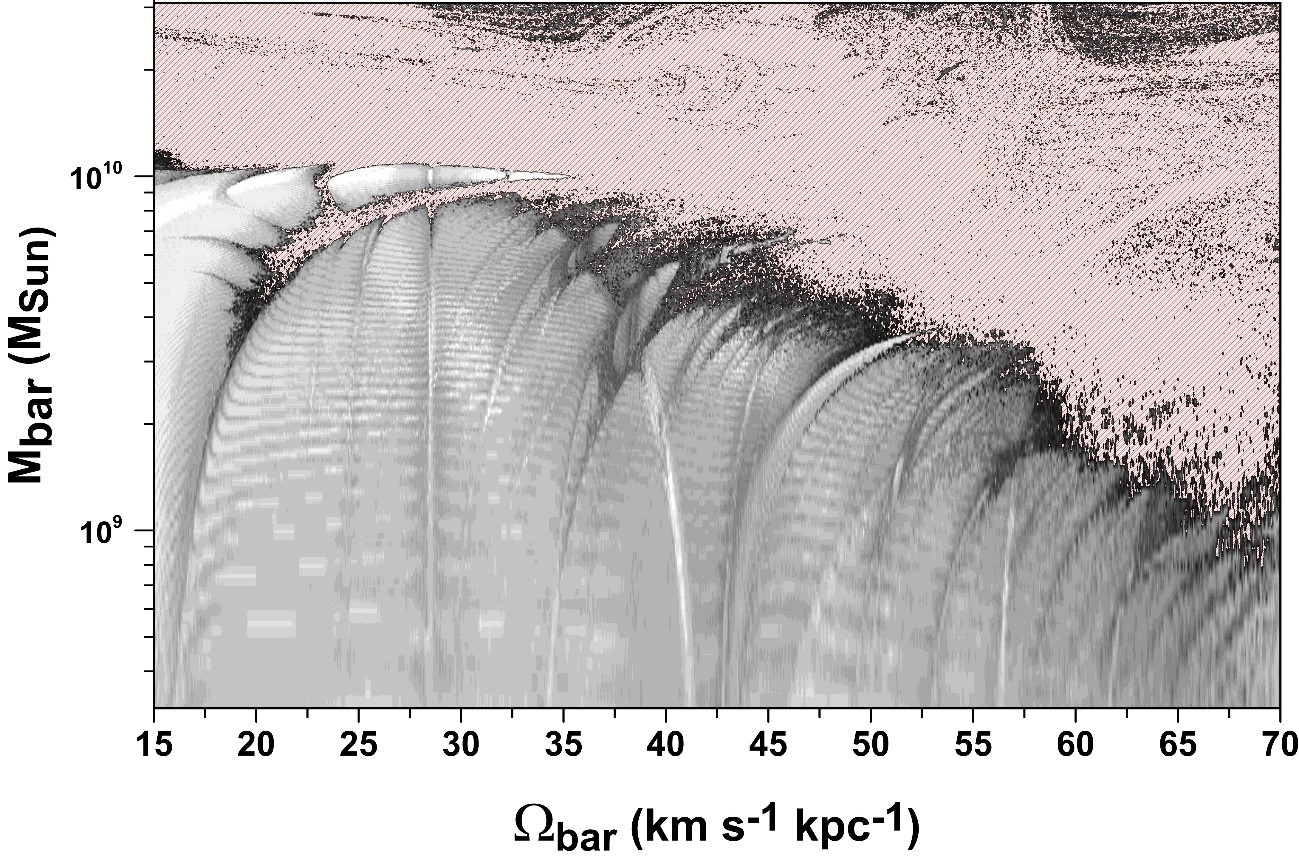,width=0.99\columnwidth ,angle=0}
\caption{Dynamical map on the parametric plane $\Omega_{\rm bar}$--$M_{\rm bar}$ for the orbit with initial conditions $R = R_{\rm bar}$, $\varphi = \gamma^0_{\rm bar}$, $p_R = 0$ and $V_\theta = V_{\rm rot} (R_{\rm bar})$, representative of the bar structure. The light grey tones represent regular orbits, while increasingly dark tones correspond to increasing instabilities and chaotic motion. The  red-hatched regions contain strongly unstable and escaping orbits. We see that, in order to have stability, one must always have $M_{\rm bar} < 10^{10} M_\odot$. This value decreases with $\Omega_{\rm bar}$. For high values of $\Omega_{\rm bar}$, the region with $M_{\rm bar} \sim 10^9 M_\odot$ is unstable.
}
\label{fig:Omega-bar-GMbar-O}
\end{center}
\end{figure}


\section{Dependence on the rotation speed of the bar}\label{sec:bar-speed}

There is enough evidence that the corotation radius of the spiral structure is close to the solar radius (see Paper II and references therein). We assume that the spiral arms rotate with the pattern speed fixed at $\Omega_{\rm sp}=28.5$\,km\,s$^{-1}$\,kpc$^{-1}$. The bar/bulge is also rotating, but its angular speed $\Omega_{\rm bar}$ is considered as a free parameter, which varies between 15 and 70\,km\,s$^{-1}$\,kpc$^{-1}$ in this section. We also vary the bar's mass from $1\times 10^8$ to $1\times 10^{11}$ solar masses; the values of the rest of the bar's parameters are fixed at those shown in Table\,\ref{tab:1}. In particular, we fix $R_{\rm bar} = 2.9$\,kpc.

In a physically plausible model, the observable structures must be stable over at least a few billion years (see discussion in Paper II). There are two physical structures present in our model which should remain stable over this period: the local arm, located near the solar radius, and the Galactic bar.
Let us first consider the local arm. Since it is associated with the spiral corotation zone (see Paper II), the stability of this zone is important for preserving the local arm structure. We analyse the stability of the spiral corotation zone as a function of the parameters $\Omega_{\rm bar}$ and $M_{\rm bar}$.  We consider in this section, as representative of this zone, the orbit of the $L_4$--centre calculated with the parameters of the basic model (see Table\,\ref{tab:1}). Perturbations due to different angular speeds of the bar structure affect the orbit of this point, which can even become chaotic.

The top of Fig.\,\ref{fig:Omega-bar-GMbar-L4} shows the dynamical map for the $L_4$--centre on the parametric plane $\Omega_{\rm bar}$--$M_{\rm bar}$. The increasingly dark tones indicate the appearance of dynamical instabilities and strong chaotic motions (red hatched regions). Regions of chaotic motion are associated, for lower values of the bar's mass ($\sim 10^9M_\odot$), with resonance regions generated by the interaction between the bar's Lindblad resonances and the spiral corotation zone. The spiral corotation zone is stable for bar's masses up to $10^{10} M_\odot$, at least in regions far from the main resonances.

The bottom part of Fig.\,\ref{fig:Omega-bar-GMbar-L4} shows the dynamical power spectrum for the $L_4$--orbit calculated along $\Omega_{\rm bar}$--axis, for fixed $M_{\rm bar}=1\times 10^9M_\odot$. We clearly see the regions in which the bar's main Lindblad resonances cross the spiral corotation zone. For instance, for $\Omega_{\rm bar}\approx\Omega_{\rm sp}$, we see simultaneous features in the spiral and bar's corotation resonances, while, for $\Omega_{\rm bar}\approx 47$\,km\,s$^{-1}$\,kpc$^{-1}$, we see simultaneous
features in the spiral corotation and the bar's OLR.
The nominal values of the bar's Lindblad resonances are denoted by vertical dashed lines.


In non-linear dynamics studies, this phenomenon is known as an overlap of resonances (see details in \citealp{LichtenbergLieberman}). It happens when  two (or more) distinct resonances are sufficiently close to each other in a phase space and, consequently, their overlap results in the appearance of widespread (large-scale) chaos. In our case, there are two distinct sources of resonances: the spiral arms rotating with the pattern speed $\Omega_p=28.5$\,km\,s$^{-1}$\,kpc$^{-1}$, and the bar rotating with the speed $\Omega_{\rm bar}$, which is varied between 15 and 70\,km\,s$^{-1}$\,kpc$^{-1}$. For the bar's corotation resonance, for instance, the condition $\Omega_p \approx \Omega_{\rm bar}$ (but not $\Omega_p\equiv\Omega_{\rm bar}$) will produce the overlap with the spiral corotation zone and, consequently, generate dynamical instabilities. This is what we observe in Fig.\,\ref{fig:Omega-bar-GMbar-L4}, where the domains (in darker tones) surrounding the nominal position of the main resonances are chaotic.

Assuming that observable objects avoid the domains of high instabilities, we can deduce the constraint on the rotation speed of the bar: its value must lie outside the zones of influence of the strong low-order resonances on the parametric plane. However, as shown in all previous sections, the situation is different when the bar's speed exactly matches the pattern speed; in this case, the spiral arms and the bar/bulge form a unique structure whose origin would still need to be explained.

Another physical structure, which must also be stable, is the bar itself. In order to quantify this stable behaviour, we associate to the bar an orbit which starts at its near extremity, with initial conditions $R = R_{\rm bar}$, $\varphi = \gamma^0_{\rm bar}$, $p_R = 0$ and $V_\theta = V_{\rm rot} (R_{\rm bar})$. We consider the stability of this orbit as an indicator of the bar's stability. The dynamical map for this orbit on the parametric plane $\Omega_{\rm bar}-M_{\rm bar}$ is shown in Fig.\,\ref{fig:Omega-bar-GMbar-O}. We see that, for all values of $\Omega_{\rm bar}$, the orbit is stable only for $M_{\rm bar} < 10^{10} M_\odot$, for low $\Omega_{\rm bar}$.  Moreover, for high $\Omega_{\rm bar}$, a bar's mass of $\sim 10^9 M_\odot$ leads to a high degree of instability for the orbit. For this range of masses, stability of this orbit imposes an upper limit of $\Omega_{\rm bar}< 50$\,km\,s$^{-1}$\,kpc$^{-1}$.


In summary, for order-of-magnitude estimates, the analysis done here is sufficient to constrain the bar's mass to $M_{\rm bar}\sim 2\times 10^9 M_\odot$ and the bar's angular velocity to $\Omega_{\rm bar} < 50$\,km\,s$^{-1}$\,kpc$^{-1}$. This leads to a situation wherein it is unlikely that the bar's OLR lies near the solar radius, since it should then have $\Omega_{\rm bar}\approx 47$\,km\,s$^{-1}$\,kpc$^{-1}$, which is close to the upper acceptable limit for $\Omega_{\rm bar}$ (see Fig.\,\ref{fig:Omega-bar-GMbar-O}).  The bar's OLR is most probably outside the solar radius.  Moreover, Fig.\,\ref{fig:Omega-bar-GMbar-L4} corroborates this conclusion. The chaotic region originated by the resonance overlap in the local corotation zone, for $\Omega_{\rm bar}\approx 47$\,km\,s$^{-1}$\,kpc$^{-1}$, would be an obstacle for the formation of the local arm (see Paper II).


\section{Possible observational evidence supporting our model}\label{sec:evidence}

\begin{figure}
\begin{center}
\epsfig{figure=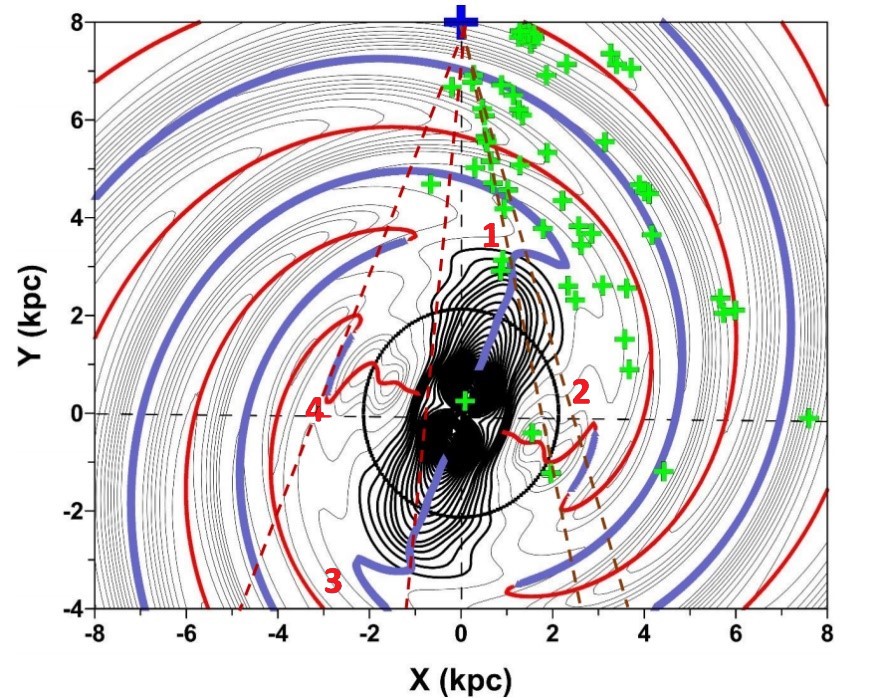,width=0.99\columnwidth ,angle=0}
\caption{
As in Fig.\,\ref{fig:U1}. The regions from \textbf{1} to \textbf{4} in the central zone of the Galaxy are defined in the text. Green crosses are masers with VLBI measurement of distance. Blue lines represent potential minima (spiral arms) and red lines potential maxima. The brown dashed lines are tangential lines passing through regions from \textbf{1} to \textbf{4}.  The circle represents the radius of the ILR of the spiral structure, which is the same as the bar if we adopt the same rotation speed of the two components.
}
\label{fig:evidences}
\end{center}
\end{figure}


\subsection{The distribution of masers in the central region of the Galaxy }\label{sec:evidence-1}

Our model of the bar predicts the existence of four regions of
complicated geometry around it: the zig-zags in the grooves which represent the minima of gravitational potential, at the extremities of the bar, and the short sectors of spiral arms, at 90$^\circ$ from the main axis. We numbered these regions from \textbf{1} to \textbf{4} in Fig.\,\ref{fig:evidences}. The Galactic longitudes of these regions can be easily determined graphically by tracing lines joining them to the position of the Sun and measuring their angle with respect to the Y axis as shown in the figure; they are, respectively, 12$^\circ$, 16$^\circ$, -5$^\circ$ (355$^\circ$) and -22$^\circ$ (338$^\circ$).

The distribution of masers with accurate positions measured with Very Long Baseline Interferometry (VLBI; see the Bessel Survey\footnote{http://bessel.vlbi-astrometry.org}) are displayed in Fig.\,\ref{fig:evidences} by green crosses; they allow us to impose restrictions on the size and orientation of the bar. These masers are associated with massive star-formation regions and molecular clouds. Observations of external barred galaxies tell us that there is no star formation inside bars \citep{James2018}. Therefore, the abrupt cut in the space density of masers at the nearest extremity of the bar tells us that the size of the bar drawn in the figure is approximately correct, confirming its length of approximately 3\,kpc.

Let us now consider the two masers observed by \cite{Sannaetal2014ApJ}, one H$_2$O maser with longitude $l=10^\circ.472$ and distance $d=8.55$\,kpc with respect to the Sun, and the other a methanol maser with $l=12^\circ.025$ and $d = 9.43$\,kpc. Both appear to coincide with the ILR, in a direction where the influence of the bar is minimal. The parallax measurements of these two masers are quite accurate, with errors of about 0.008 mas. The LSR velocities of the two sources are high (69 and 108\,km\,s$^{-1}$, respectively), which is consistent with their proximity to the Galactic centre. We know, from the experience with the local arm discussed in Paper II, that zones of resonances also harbour regions of star formation, with the presence of masers. The coincidence of the two masers with the small-sized zone of potential maximum (as indicated by the contour lines in Fig.\,\ref{fig:evidences}) seems to establish a strong restriction on the inclination of the bar, since a small rotation of the bar would destroy the coincidence, as well as a restriction on the width of the bar.

For the symmetric positive potential zone on the other side of the bar (region \textbf{4}), we did not find any report of VLBI observations of masers, but interesting data are available in the literature. \cite{greenEtal2011ApJ} investigated the distribution of methanol maser sources in the inner Galaxy, and found the most prominent tangential direction at a longitude of  $-22^\circ$ (or $338^\circ$) with a cumulation of masers with a range of velocities that reaches $-80$\,km\,s$^{-1}$. The authors attribute this feature to the Perseus arm origin. We note that the segment of arm in region \textbf{4} (with negative potential, indicated in blue) is not the origin of the Perseus arm, in our model. However, we can see from Fig.\,\ref{fig:evidences} that it is situated in the direct prolongation of the Perseus arm, and could be interpreted as being part of it. The observations of Green et al. reinforce our interpretation that region \textbf{4}, situated at a longitude of $-22^\circ$, is a region containing methanol masers.

The only arms present in our model are the four spiral arms already discussed in Paper II. In our interpretation of the nature of the arms, there are no “expanding” arms, as often mentioned in the literature \citep[e.g.][]{Sannaetal2009ApJ}. The arms rotate with the general pattern speed, without moving, one with respect to the others. We believe that the observed anomalous velocities are due to the flow of gas along the arms. Due to the pitch angle of the arms and of their zig-zags, velocity components in the direction of the Sun may appear. Detailed models of the zig-zags are left for future work.

\begin{figure}
\begin{center}
\epsfig{figure=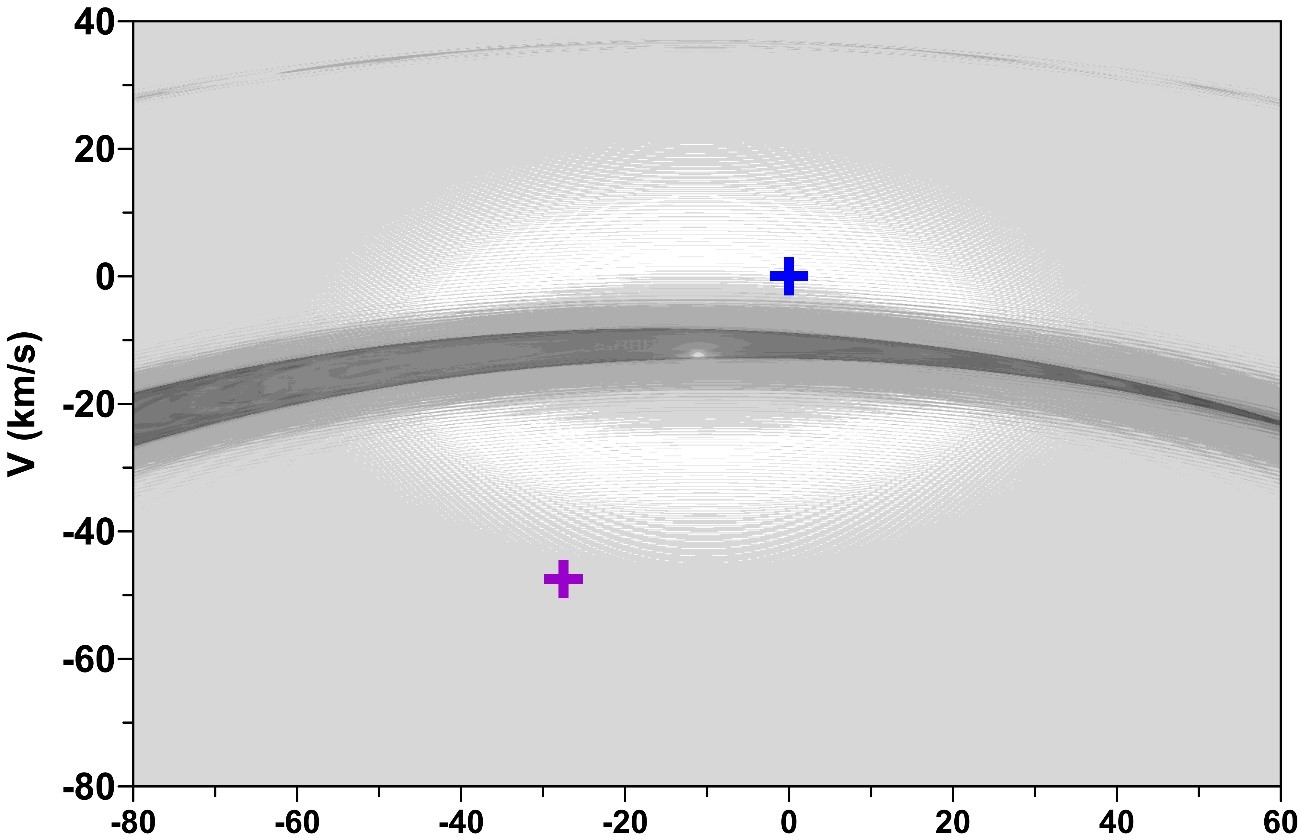,width=0.99\columnwidth ,angle=0}
\epsfig{figure=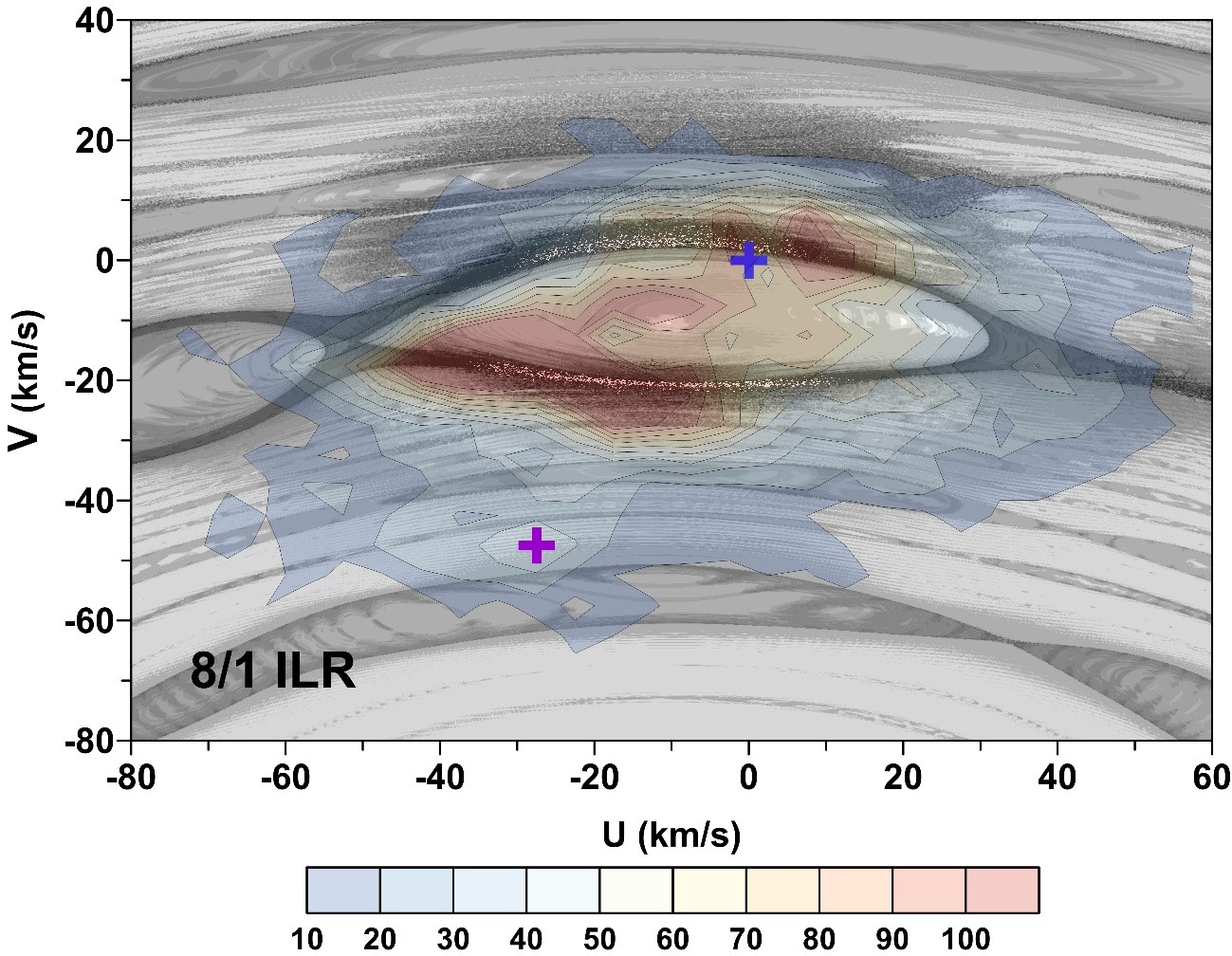,width=0.99\columnwidth ,angle=0}
\caption{Top: Dynamical map on the heliocentric $U$--$V$ plane. Only perturbations due to the central bar/bulge are accounted. The light gray tones represent regular orbits, while increasingly dark tones correspond to increasing instabilities and chaotic motion. The position of the Sun and of the density peak of Hercules are shown by blue and magenta crosses, respectively.
Bottom: Same as on the top graph, except including perturbations due to the spiral arms; the values of the parameters are taken from Table\,\ref{tab:1}. The iso-density contours of the density of stars are superimposed and the color bar associates the colors to the density values (in units of number of objects per a bin of $4\times4$\,km\,s$^{-1}$).
}
\label{fig:plane-U-V}
\end{center}
\end{figure}


\subsection{Main moving groups and Hercules stream}\label{sec:evidence-2}

Many efforts have been dedicated to assessing the origin of the Hercules stellar stream, detected in the velocity space of the solar neighbourhood. \cite{BensbyEtalApJL2007} analysed age and abundance distributions of stars in the Hercules stream and found them to be similar to the trends in the thin and thick discs, thus concluding that the stream is a mixture of thin and thick disc stars. 

Since the work from \cite{dehnen2000AJ}, the origin of the Hercules stream is believed to be due to the dynamical effects of a fast rotating bar, whose OLR would be placed in the vicinity of the solar radius. For this, the pattern speed of the bar should be 1.85 times the angular velocity at the Sun's position. A similar result was obtained recently by \cite{MonariEtalMNRAS2017}, by combining the catalogues from Gaia DR1 and LAMOST and verifying that the variation of the position of Hercules in velocity space as a function of Galactic radius matched the ones predicted by fast bar models. The authors also point out the contradictory results that come from photometric and spectroscopic stellar surveys and the gas kinematics in the inner Galaxy, which favour a slowly rotating bar. A reconciliation of the Hercules stream with a slow bar model was achieved by \cite{PerezVillegasEtal2017}. The authors propose the Hercules stream as being made of stars orbiting the Lagrange points of the bar with a pattern speed of 39\,km\,s$^{-1}$\,kpc$^{-1}$; the stars move outward from the bar's corotation radius and visit the solar neighbourhood.

\cite{quillen2003AJ} studied chaos caused by resonance overlap in a model putting the solar neighbourhood near the ILR of a two-armed spiral pattern, or near the 4/1 ILR in the case of a four-armed pattern, and verified that the stellar orbits supporting the spiral  structure and those oscillating with the bar are disrupted near the bar's OLR; the spiral structure refined the boundaries of the Hercules stream in the local velocity space. \cite{FuxAA2001} also interpreted the Hercules stream as an overdensity of chaotic orbits due to the rotating bar \citep[e.g.][]{FamaeyEtalAA2005}.

From the aforementioned results, we would not expect that our bar model, with a pattern speed of 28.5\,km\,s$^{-1}$\, kpc$^{-1}$, placing the bar's corotation in the vicinity of the solar radius, could generate a bimodal feature in the local velocity space in the same way as the cited studies obtained putting the bar's OLR near the Sun's position. Indeed, we verified, from a dynamical map of the solar neighbourhood on the $U$--$V$ plane, that a model accounting solely for the bar and bulge perturbations (with the values of the parameters taken from Table\,\ref{tab:1}) does not produce a Hercules-stream-like feature. In the top panel of Fig.\,\ref{fig:plane-U-V}, we show the dynamical map of the $U$--$V$ plane calculated for such a model.

On the other hand, a model accounting for the spiral arms perturbation produces more interesting features in the dynamical map of the $U$--$V$ plane. The bottom panel of Fig.\,\ref{fig:plane-U-V} shows the iso-density contours of stars in the observed $U$--$V$ plane of the solar neighbourhood, taken from data of the Geneva-Copenhagen survey catalogue \citep{HolmbergEtal2009}, superposed to the dynamical map of the modelled $U$--$V$ plane. In this case, the model includes the perturbation from the spiral arms and from the bar/bulge structure.

The main moving groups (Pleiades, Hyades, Sirius, and Coma Berenices) lie, approximately, inside the spiral corotation zone. We see a clear correlation between the observed structures and this resonance (in the central region of the plane). Apart from the relationship between the main moving groups  and the corotation zone, we focus next on the Hercules stream and the chains of resonances seen in Fig.\,\ref{fig:plane-U-V}.

The more prominent island of stability in the bottom part of the dynamical map, with $V < -50$\,km\,s$^{-1}$, is associated with the 8/1 ILR of the spiral pattern. Here we propose this resonance as a likely dynamical origin of the Hercules stream, given the proximity between the extensions of these two features in the $U$--$V$ plane. At first sight, a connection between these structures is not particularly evident; this is due to the fact that the orbits for the construction of the dynamical maps on the $U$--$V$ plane in Fig.\,\ref{fig:plane-U-V} were integrated fixing the initial conditions of the test-particle at $R = 8$\,kpc and $\varphi=90^{\circ}$, and varying the $U$ and $V$ velocity values. As a consequence,  these conditions influence the aspects and positions of the chains of resonances in the $U$--$V$ plane.
In fact, if we fix the initial $U$--value of the orbits at $-28$\,km\,s$^{-1}$, the 8/1 resonance island is seen to extend from $V = -60$\,km\,s$^{-1}$ to $V = -40$\,km\,s$^{-1}$ (varying R  between 7.7 and 8.3 kpc), thus better matching the position of the Hercules stream in the observed U–V plane. Also, small displacements of the initial radius, for example to R = 7.8 or 7.9 kpc,  move the whole 8/1 resonance upwards and produce good matches between this resonance and the Hercules stream. We also verified that several orbital trajectories of stars in the Hercules stream present a radial oscillation compatible with being inside a 8/1 resonance or quasi-resonance with the spiral pattern. Minor contributions from the 9/1 and 10/1 resonances may also be expected.

We are the first ones to relate the Hercules stream to the 8/1 ILR and higher-order resonances between a four-armed spiral pattern structure and the stars in the disc, with the solar neighbourhood in the vicinity of the corotation radius. A thorough analysis of this association, with simulations of the local $U$--$V$ plane produced by stellar orbits under the perturbed potential model, as well as the relation between the main moving groups and the corotation zone, is left for future work.

\section{Discussion and conclusions}\label{sec:conclus}

We have developed a model for the equatorial plane of the Galaxy, which includes the central bar/bulge structure and the spiral arms. The simultaneous perturbations of the two components on the stellar motion are of fundamental importance to constrain the range of acceptable Galactic parameters and to understand the nature of the resonances that are observed in the solar neighbourhood. The model of spiral arms was already presented in Paper I, and most of the effort was now directed to the bar/bulge model. 

Our basic model considers a bar as an elongated ellipsoid with a semimajor axis of 2.9\,kpc and flattening of 0.7, oriented at $22^\circ.5$ from the Sun-Galactic centre direction. Two of the spiral arms are connected to the extremities of the bar. The total mass of the bar is 10$^9 M_\odot$, which is distributed uniformly inside it. The model includes also a spheroidal bulge with a total mass of 10$^{10} M_\odot$, whose perturbations are axisymmetric in the equatorial plane and produce no effect on the resonances. In the basic model, the bar is supposed to have the same rotation speed of the spiral arms, 28.5\,km\,s$^{-1}$\,kpc$^{-1}$, so that the corotation radii of the bar and of the spiral arms coincide. The bar produces a minimum in the gravitational potential, which almost coincides, in the azimuthal direction, with the zone of corotation produced by the spiral arms. However, for a bar with mass equal to $10^9M_\odot$, the corotation zone is not affected, since the magnitude of the perturbation of the bar, at the solar radius, is much smaller than that of the spiral arms.

We performed tests increasing the mass of the bar, and noticed that the remaining $L_4$--centre of corotation becomes displaced towards larger values of the azimuthal angle $\varphi$ and Galactic radius $R$. An upper limit of about $5\times 10^9M_\odot$ can be set, when this displacement becomes incompatible with the observed position of the local arm, which is supposedly associated with the $L_4-$centre of the corotation zone (see Paper II). Pointing to the same restriction, the orbit of the Sun becomes chaotic if we adopt a mass of the bar of $5\times 10^9M_\odot$, keeping the other parameters the same.

Variations around the parameters of the basic model show that the stability of the local arm does not impose any strict range for the bar's angular velocity $\Omega_{\rm bar}$, except for a few regions of influence of the main resonances of the spiral pattern. On the other hand, the stability of the bar itself is also a requirement our model must satisfy. Taking the near extremity of the bar as representative for its stability, we find that the bar's mass must be of order of a few $M_{\rm bar} \sim 10^9M_\odot$, and the bar's angular velocity is restricted by $\Omega_{\rm bar} < 50$\,km\,s$^{-1}$\,kpc$^{-1}$ (approximately). Our choice of a common  pattern speed for the arms and the bar in our basic model is only justified because it seems to be less arbitrary than any other choice, and a stable connection between the arms and the bar is easier to model and is not in  conflict with the observations.

Although we have not attempted to fine tune  any of the parameters, it seems that the basic set of parameters generates a model which best fits the observed  size of the bar, and produces no deformation of the spiral arms situated between the bar and the Sun, in agreement with the maps of the Galaxy by \cite{houHan2014AA}. This model does not disturb the zone of corotation situated near the Sun and gives the best non-chaotic area around it. The ratio between corotation and bar radius is 2.7, which would be too high for an early type galaxy, but is perfectly acceptable for an SBbc galaxy, such as the Milky Way \citep{ElmegreenElmegreen1989ApJ}.

We showed that the size and the orientation of the bar are well restricted by the position of maser sources which have distances given by VLBI measurements. We suggest here that the longer bar reaching 4.5\,kpc observed by \cite{Lopez_CorredoiraEtal2007} may be due to the zone of transition between the spiral arms and the bar, which could appear as a prolongation of the bar.

Our results suggest that the main moving groups of the solar neighbourhood (Pleiades, Hyades, Sirius, and Coma Berenices) belong to the spiral corotation zone. Moreover, we show that the Hercules stream, constituted of stars of the solar neighbourhood expanding towards the anti-centre and in counter-rotation direction with respect to the LSR, is close to the  zone of stability of the 8/1 resonance of the spiral arms. We suggest that this is a better explanation for the Hercules stream than the  OLR of the bar, which has been proposed by several authors. Indeed, that hypothesis would require a pattern speed of 47\,km\,s$^{-1}$\,kpc$^{-1}$, for which there is no independent observational evidence. Also, this value is on the edge of the acceptable range of parameters dynamically constrained by our model, so it seems unlikely that the bar's OLR lies close to the solar radius. Moreover,  if the bar's OLR is close to the solar radius, resonance overlap with the spiral corotation zone will generate a wide chaotic region which would compromise the stability of the local arm structure.

The present model, due to its nature, can be used to predict the $U$--$V$ velocity distribution for farther distances from the Sun, as done in Bovy (2010). The predictions could be compared with the forthcoming Gaia DR 2, in order to test whether our model is able to reproduce the velocity distribution observed in the extended solar neighbourhood. This is left for future work.


\begin{acknowledgements}

We acknowledge Dr. Angeles P{\'e}rez-Villegas for critical reading of the manuscript and useful suggestions. This work was supported by the São Paulo State Science Foundation, FAPESP, and the Brazilian National Research Council, CNPq. R.S.S.V. acknowledges FAPESP grant 2015/10577-9. This work has made use of the facilities of the Laboratory of Astroinformatics (IAG/USP, NAT/Unicsul), whose purchase was made possible by FAPESP (grant 2009/54006-4) and the INCT-A. We acknowledge the anonymous referee for the detailed review and for the helpful suggestions which allowed us to improve the manuscript.

\end{acknowledgements}




\begin{thebibliography}{73}
\expandafter\ifx\csname natexlab\endcsname\relax\def\natexlab#1{#1}\fi

\bibitem[Ablimit \& Zhao(2017)]{Ablimit_Zhao2017} Ablimit, I., \& Zhao, G.\ 2017, \apj, 846, 10

\bibitem[{{Allen} \& {Santill\'an}(1991)}]{Allen_Santillan1991}
{Allen}, C. \& {Santill\'an}, A. 1991, Rev. Mexicana Astron. Astrofis., 22, 255

\bibitem[Antoja {et al.}(2008)]{antoja2008} {Antoja}, T., {Figueras}, F., {Fern{\'a}ndez}, D. \& Torra, J. 2008, \aap, 490,135

\bibitem[Antoja {et al.}(2009)]{antoja2009} {Antoja}, T., {Valenzuela}, O., {Pichardo}, B., {Moreno}, E.,   {Figueras}, F. \& {Fern{\'a}ndez}, D. 2009, \apjl, 700, L78

\bibitem[Antoja {et al.}(2011)]{antojaEtal2011MNRAS} {Antoja}, T., {Figueras}, F., {Romero-G{\'o}mez}, M., {et~al.} 2011, \mnras,
  418, 1423


\bibitem[Antoja {et al.}(2012)]{antojaEtal2012MNRAS} {Antoja}, T., {Helmi}, A., {Bienayme}, O., {et~al.}  2012, \mnras,
  426, L1

\bibitem[Antoja et al.(2014)]{AntojaEtalAA2014} Antoja, T., Helmi, A., Dehnen, W., et al.\ 2014, \aap, 563, A60

\bibitem[Barros et al.(2016)]{BarrosEtal2016AA} Barros, D.~A., L{\'e}pine, J.~R.~D., \& Dias, W.~S.\ 2016, \aap, 593, A108

\bibitem[Bensby et al.(2007)]{BensbyEtalApJL2007} Bensby, T., Oey, M.~S., Feltzing, S., \& Gustafsson, B.\ 2007, \apjl, 655, L89

\bibitem[Bienaym{\'e}(2017)]{Bienayme2017} Bienaym{\'e}, O.\ 2017, arXiv:1710.07065

\bibitem[Bland-Hawthorn \& Gerhard (2016)]{Bland-Hawthor2016} {Bland-Hawthorn}, J. \& {Gerhard}, O. 2016, \araa, 54, 529

\bibitem[Bobylev et al.(2014)]{BobylevEtal2014} Bobylev, V.~V., Mosenkov, A.~V., Bajkova, A.~T., \& Gontcharov, G.~A.\ 2014, Astronomy Letters, 40, 86
\bibitem[Bovy (2010)]{Bovyl2010} Bovy, J. 2010, \apj, 725, 1676

\bibitem[{{Chandrasekhar}(1969) }]{chandrasekhar1987} {Chandrasekhar},S., 1987,
 in Ellipsoidal Figures of Equilibrium, Yale Univ. Press., New Haven

\bibitem[{{Clemens}(1985)}]{clemens1985ApJ}{Clemens}, D.~P. 1985, \apj, 295, 422

\bibitem[{{Combes} \& {Elmegreen}(1993)}]{CombesElmegreen93}
{Combes}, F. \& {Elmegreen}, B.~G. 1993, \aap, 271, 391

\bibitem[{{Contopoulos}(1980)}]{Contopoulos1980}{Contopoulos}, G. 1980, \aap, 81, 198

\bibitem[{{Dehnen}(2000)}]{dehnen2000AJ}{Dehnen}, W. 2000, \aj, 119, 800

\bibitem[{{Drimmel} \& {Spergel}(2001)}]{drimmelSpergel2001ApJ}{Drimmel}, R. \& {Spergel}, D.~N. 2001, \apj, 556, 181

\bibitem[{{Duboshin}(1968) }]{Duboshin1968} {Duboshin}, G.~N. 1968,
 in Celestial Mechanics. Basic Problems and Methods, "Nauka", Moskow

\bibitem[Eggen(1996)]{Eggen1996} Eggen, O.~J.\ 1996, \aj, 112, 1595

\bibitem[{{Elmegreen} \& {Elmegreen}(1989)}]{ElmegreenElmegreen1989ApJ}
{Elmegreen}, B.G. \& {Elmegreen}, D.M. 1989, \apj, 342, 677

\bibitem[Famaey et al.(2005)]{FamaeyEtalAA2005} Famaey, B., Jorissen, A., Luri, X., et al.\ 2005, \aap, 430, 165

\bibitem[{{Ferraz-Mello} {et~al.}(2005){Ferraz-Mello}, {Michtchenko},
  {Beaug{\'e}}, \& {Callegari}}]{ferrazmeloMichtchenkoEtal2005LNP}
{Ferraz-Mello}, S., {Michtchenko}, T.~A., {Beaug{\'e}}, C., \& {Callegari}, N.
  2005, in Lecture Notes in Physics, Berlin Springer Verlag, Vol. 683, Chaos
  and Stability in Planetary Systems, ed. R.~{Dvorak}, F.~{Freistetter}, \&
  J.~{Kurths}, 219--271

\bibitem[{{Fich} {et~al.}(1989){Fich}, {Blitz}, \&
  {Stark}}]{fichBlitzStark1989ApJ}
{Fich}, M., {Blitz}, L., \& {Stark}, A.~A. 1989, ApJ, 342, 272

\bibitem[Fux(2001)]{FuxAA2001} Fux, R.\ 2001, \aap, 373, 511

\bibitem[{{Georgelin} \& {Georgelin}(1976)}]{georgelinGeorgelin1976AA}
{Georgelin}, Y.~M. \& {Georgelin}, Y.~P. 1976, \aap, 49, 57

\bibitem[Gerhard(2017)]{Gerhard2017} Gerhard, O.\ 2017, arXiv:1710.01544

\bibitem[Gnedin et al.(2010)]{GnedinEtalApJL2010} Gnedin, O.~Y., Brown, W.~R., Geller, M.~J., \& Kenyon, S.~J.\ 2010, \apjl, 720, L108

\bibitem[{{Green} {et~al.}(2011){Green},{Caswell},{McClure-Griffiths},
{Avison}, {Breen}, {Burton}, {Ellingsen}, {Fuller}, {Gray}, {Pestalozzi},
{Thompson}, \& {Voronkov}}]{greenEtal2011ApJ}{Green}, J.A., {Caswell}, J.L.,
{McClure-Griffiths},N.M.,{Avison}, A.,{Breen}, S.L.,{Burton}, M.G., {Ellingsen},
S.P.,{Fuller}, G.A., {Gray}, M.D., {Pestalozzi}, M.,{Thompson}, M.A.,
 \& {Voronkov}, 2011, ApJ 733, 27

\bibitem[Holmberg et al.(2009)]{HolmbergEtal2009} Holmberg, J., Nordstrom, B., \& Andersen, J.\ 2009, VizieR Online Data Catalog, 5130,

\bibitem[{{Hou} \& {Han}(2014)}]{houHan2014AA}{Hou}, L.~G. \& {Han}, J.~L. 2014, \aap, 569, A125

\bibitem[{{James} \& {Percival}(2018)}]{James2018}{James}, P.~A. \& {Percival}, S.~M. 2018, \mnras, 474, 3101

\bibitem[{{Junqueira} {et~al.}(2013){Junqueira}, {L{\'e}pine}, {Braga}, \&
  {Barros}}]{junqueiraEtal2013AA}
{Junqueira}, T.~C., {L{\'e}pine}, J.~R.~D., {Braga}, C.~A.~S., \& {Barros},
  D.~A. 2013, \aap, 550, A91

\bibitem[{{L{\'e}pine} \& {Leroy}(2000)}]{Lepine_Leroy2000}
{L{\'e}pine}, J.~R.~D. \& {Leroy}, P. 2000, MNRAS, 313, 263

\bibitem[{{L{\'e}pine} {et~al.}(2017){L{\'e}pine}, {Michtchenko}, {Barros}, \& {Vieira}
  }]{lepineEtal2017ApJ} {L{\'e}pine}, J.~R.~D., {Michtchenko}, T.~A., {Barros}, D.~A., {Vieira}, R.~S.~S. 2017, \apj,
  843:48 (Paper II)

\bibitem[{{Lichtenberg \& Lieberman}(1992) {Lichtenberg}, {Lieberman} }]{LichtenbergLieberman}
{Lichtenberg}, A.~J., \& {Lieberman}, M.~A. 1992, in Regular and Chaotic Dinamics, Second Edition, Applied Mathematical Sciences 38, Springer

\bibitem[{{{\L}okas}(2016)}]{Lokas2016ApJ}
{{\L}okas}, E.~L. 2016, \apjl, 830, L20

\bibitem[L{\'o}pez-Corredoira et al.(2007)]{Lopez_CorredoiraEtal2007} L{\'o}pez-Corredoira, M., Cabrera-Lavers, A., Mahoney, T.~J., et al.\ 2007, \aj, 133, 154

\bibitem[Meidt et al. (2014)]{meidt2014} {Meidt}, S.~E., {Schinnerer}, E., {van de Ven}, G., {et al.} 2014, \apj, 788, 144

\bibitem[{{Michtchenko} {et~al.}(2002){Michtchenko}, {Lazzaro}, {Ferraz-Mello},
  \& {Roig}}]{michtchenkoEtal2002Icar}
{Michtchenko}, T.~A., {Lazzaro}, D., {Ferraz-Mello}, S., \& {Roig}, F. 2002,
  \icarus, 158, 343

\bibitem[{{Michtchenko} {et~al.}(2017){Michtchenko}, {Vieira}, {Barros}, \& {L{\'e}pine}}]{MichtchenkoEtal2017ApJ} {Michtchenko}, T.~A., {Vieira}, R.~S.~S.,  {Barros}, D.~A., {L{\'e}pine}  J.~R.~D. 2017, \aap,
  597, A39 (Paper I)

\bibitem[Mihalas \& Binney(1981)]{Mihalas_Binney1981} Mihalas, D., \& Binney, J.\ 1981, San Francisco, CA, W.~H.~Freeman and Co., 1981.~608 p.,

\bibitem[Miller \& Smith (1979)]{MillerSmith1979} {Miller}, R.~H. \& {Smith}, B.~F. 1979, \apj, 227,785

\bibitem[Monari et al.(2017)]{MonariEtalMNRAS2017} Monari, G., Kawata, D., Hunt, J.~A.~S., \& Famaey, B.\ 2017, \mnras, 466, L113

\bibitem[P{\'e}rez-Villegas et al.(2017)]{PerezVillegasEtal2017} P{\'e}rez-Villegas, A., Portail, M., Wegg, C., \& Gerhard, O.\ 2017, \apjl, 840, L2

\bibitem[{{Pichardo} {et~al.}(2004){Pichardo}, {Martos}, \&
  {Moreno}}]{pichardoMartosMoreno2004ApJ}
{Pichardo}, B., {Martos}, M., \& {Moreno}, E. 2004, \apj, 609, 144

\bibitem[{{Portail} {et~al.}(2017){Portail}, {Gerhard}, {Wegg}, \& {Ness}}]{Portail2017}
{Portail}, M., {Gerhard}, O., {Wegg}, C., \& {Ness}, M. 2017, \mnras, 465 1621

\bibitem[{Powell \& Percival(1979)}]{powell1979}
Powell, G.~E. \& Percival, I.~C. 1979, Journal of Physics A: General Physics,
  12, 2053

\bibitem[{{Quillen}(2003)}]{quillen2003AJ}
{Quillen}, A.~C. 2003, \aj, 125, 785


\bibitem[{{Reid} {et~al.}(2014){Reid}, {Menten}, {Brunthaler}, {Zheng}, {Dame},
  {Xu}, {Wu}, {Zhang}, {Sanna}, {Sato}, {Hachisuka}, {Choi}, {Immer},
  {Moscadelli}, {Rygl}, \& {Bartkiewicz}}]{reidEtal2014ApJ}
{Reid}, M.~J., {Menten}, K.~M., {Brunthaler}, A., {et~al.} 2014, \apj, 783, 130

\bibitem[{{Russeil}(2003)}]{russeil2003AA}
{Russeil}, D. 2003, \aap, 397, 133

\bibitem[Salo et al. (2015)]{Salo2015}{Salo}, H., {Laurikainen}, E., {Laine}, J. {et~al.} 2015, \apjs, 219, 4

\bibitem[Sanna et al.(2009)]{Sannaetal2009ApJ} Sanna, A., Reid, M.~J., Moscadelli, L., et al. 2009, \apj, 706, 464

\bibitem[Sanna et~al.(2014)]{Sannaetal2014ApJ}
{Sanna}, A., {Reid}, M.~J., {Dame}, T.M., {et~al.} 2014, \apj, 781, 108

\bibitem[{{Skuljan} {et~al.}(1999){Skuljan}, {Hearnshaw}, \&
  {Cottrell}}]{skuljanEtal1999MNRAS}
{Skuljan}, J., {Hearnshaw}, J.~B., \& {Cottrell}, P.~L. 1999, \mnras, 308, 731

\bibitem[{{Sofue}(2013) {Sofue}}]{sofue2013PSSS}
{Sofue}, Y., 2013, in Planets, Stars and Stellar Systems, Springer,
 Berlin 2013, Vol. 5, ed. G. {Gilmore}, Chap 19

\bibitem[{{Sofue} \& {Nakanishi}(2016)}]{sofueNakanishi2016PASJ}
{Sofue}, Y., \& {Nakanishi}, H. 2016, \pasj, 68, 63

\bibitem[{{Sormani} {et~al.}(2015){Sormani}, {Binney}, \&
  {Magorrian}}]{SormaniBinney2015}
{Sormani}, M.~C., {Binney}, J., \& {Magorrian}, J. 2015, \mnras, 451, 3437

\bibitem[{{Vall{\'e}e}(2013)}]{vallee2013IJAA}
{Vall{\'e}e}, J.~P. 2013, International Journal of Astronomy and Astrophysics,
  3, 20

\bibitem[{{Wegg} {et~al.}(2015){Wegg}, {Gerhard}, \&
  {Portail}}]{Wegg2015}
{Wegg}, C., {Gerhard}, O., \& {Portail}, M. 2015, \mnras, 450, 4050


\end{thebibliography}



\appendix

\section{Shape and potential of a homogeneous ellipsoid of rotation}\label{app-A}

Let the surface of an ellipsoid be given by the equation
\begin{equation}\label{elipse}
  \frac{x^{2}}{a^2}+\frac{y^{2}}{b^2}+\frac{z^{2}}{c^2}=1.
\end{equation}
The ellipsoid is centred at the origin of the reference frame and its semi-axes, $a$, $b$ and $c$, are aligned to the axes $x$, $y$ and $z$, respectively. The gravitational potential of the homogeneous ellipsoid (\ref{elipse}) with the mass $M$ at the point $P(x^*,y^*,z^*)$ has the form
\begin{equation}
\label{eq:integral}
\begin{array}{rcl}
\Phi(P)&=&-\frac{3}{4}GM\int_{\lambda}^{\infty}
{\big( 1-\frac{x^{*2}}{a^2+s}-\frac{y^{*2}}{b^2+s}-\frac{z^{*2}}{c^2+s}\big )\frac{ds}{F_s}}, \\
 & & \\
F_s&=&\sqrt{(a^2+s)(b^2+s)(c^2+s)},
\end{array}
\end{equation}
where  $G$ is the universal gravitational constant; the expression above is known as \emph{Dirichlet integral formula}. The lower limit $\lambda$ is equal to zero for the calculation of the potential in the interior of the ellipsoid.  At an arbitrary point outside the body, $\lambda$ is the positive root of the equation
\begin{equation}\label{eq:root}
  \frac{x^{*2}}{a^2+\lambda}+\frac{y^{*2}}{b^2+\lambda}+\frac{z^{*2}}{c^2+\lambda}=1\,.
\end{equation}

In the case of a homogeneous ellipsoid of revolution, the potential (\ref{eq:integral}) can be written as follows.
\begin{equation}\label{eq:generic}
\Phi(P) = -\frac{3}{2}G\,M\big[U_0(\zeta)+U_1(\zeta)\,x^{*2}+U_2(\zeta)\,y^{*2}+U_3(\zeta)\,z^{*2}\big],
\end{equation}
where the coefficients $U_0(\zeta)$, $U_1(\zeta)$, $U_2(\zeta)$ and $U_3(\zeta)$ are analytic functions of the variable $\zeta$ given as
\begin{equation}\label{zeta}
\zeta^2 = \frac{a^2-c^2}{c^2+\lambda}.
\end{equation}
We reiterate that, at an external point, $\lambda$ is calculated as the positive root of the quadratic equation (\ref{eq:root}). At a point in the interior of the ellipsoid, $\lambda$ is always zero, allowing us to introduce a constant $\zeta_0$ as
\begin{equation}\label{eq:zeta0}
\zeta_0^2 = \frac{a^2-c^2}{c^2}.
\end{equation}
For an elongated ellipsoid of rotation, $a>b=c$, the constant $\zeta_0$ defines the equatorial eccentricity, while, for an oblate ellipsoid of rotation, $a=b>c$, $\zeta_0$ defines the polar eccentricity. For a sphere, the eccentricity is equal to zero, and it increases with increasing flattening of the spheroid.

For an elongated ellipsoid of rotation with $a>b=c$, the coefficients $U_i(\zeta)$ ($i=0, 1, 2, 3$) in (\ref{eq:generic}) are defined in closed form as:
\begin{eqnarray}\label{eq:elong-coeffic}
  c\,U_0(\zeta)  &=& \frac{1}{\zeta_0}\ln\left(\sqrt{1+\zeta^2}+\zeta\right), \\
  c^3\,U_1(\zeta)&=&\frac{1}{\zeta_0^3}\left[\frac{\zeta}{\sqrt{1+\zeta^2}}-\ln\left(\sqrt{1+\zeta^2}+\zeta\right)\right], \\
  c^3\,U_2(\zeta)&=&c^3\,U_3(\zeta)=\frac{1}{2\zeta_0^3}\left[\ln\left(\sqrt{1+\zeta^2}+\zeta\right)- \zeta\sqrt{1+\zeta^2}\right],
\label{eq:elong-coeffic1}\end{eqnarray}

For an oblate ellipsoid of rotation with $a=b>c$, the coefficients $U_i(\zeta)$ ($i=0, 1, 2, 3$) in (\ref{eq:generic}) are defined in closed form as:

\begin{eqnarray}\label{eq:oblate-coeffic}
  c\,U_0(\zeta)  &=& \frac{1}{\zeta_0}\arctan\left(\zeta\right), \\
  c^3\,U_1(\zeta)&=&c^3\,U_2(\zeta)=\frac{1}{2\zeta_0^3}\left[\frac{\zeta}{1+\zeta^2}-\arctan\left(\zeta\right)\right], \\
  c^3\,U_3(\zeta)&=& \frac{1}{\zeta_0^3}\left[\arctan\left(\zeta\right) - \zeta\right].
\label{eq:oblate-coeffic1}\end{eqnarray}

Finally, the potential of a homogeneous sphere with mass $M$ and radius $R_{\rm sph}$, which is a particular case of an ellipsoid of rotation with $\zeta_0=0$, is written as

\begin{equation}
\Phi_{\rm sph}(R)=-GM\times\left\{
\begin{array}{cc}
(3R_{\rm sph}^2-R^2)/2R_{\rm sph}^3, & \mbox{for}\,\, R\leq R_{\rm sph},\\
1/R,              & \mbox{for}\,\, R\geq R_{\rm{sph}}.
\end{array}
\right.
\label{eq:sphere}
\end{equation}

\section{Spectral analysis method: dynamical maps and dynamical power spectra}\label{app-B}

The Spectral Analysis Method  is a powerful method in the study of the dynamical stability of an orbit (for details, see \citealt{michtchenkoEtal2002Icar,ferrazmeloMichtchenkoEtal2005LNP}, and an application in the context of galactic dynamics in \citealt{MichtchenkoEtal2017ApJ}). The method allows to distinguish between regular and chaotic motions of dynamical systems and is based on the well-known features of power spectra \citep[plot of the amplitude of the Fourier transform of a time series against frequency, see][]{powell1979}. It involves two main steps. The first step is the numerical integration of the equations of motion defined by the full Hamiltonian (\ref{eq:eq1}). The second step consists of the spectral analysis of the output of the numerical integrations. The time series giving the variation of stellar orbital elements (e.g. the canonical phase-space coordinates) are Fourier-transformed using a standard fast Fourier transform (FFT) algorithm and the main oscillation modes are identified. For more details about these methods, see \cite{MichtchenkoEtal2017ApJ}.

\subsection{Dynamical maps on representative planes}\label{app-B2}

The power spectrum of a time series presents peaks corresponding to the main frequencies of the orbit. Regular orbits are quasi-periodic and have few frequency peaks, given by the two independent frequencies, their harmonics and linear combinations. The amplitude of these peaks, however, drops abruptly when we go to high values. Therefore, their power spectra present only few significant frequency peaks.

On the other hand, chaotic orbits are not confined to an invariant torus; they span a region with higher dimensionality than that of the invariant tori. In practice, this means that their power spectra present a quasi-continuum of frequencies, all of them with comparable magnitudes. Therefore, the number of significant frequencies (defined here as those with amplitude higher than 5\% of the largest peak in the spectrum) is a quantifier of chaos. This number is called \textit{spectral number $N$}; small values of $N$ indicate regular motion, while large values correspond to the onset of chaos. The spectral number $N$ also depends  on the integration time span; the chosen total integration time should be  large enough to  allow  chaos generated by resonances to be noticeable. It is worth noting that the method is robust against small variations of the minimum peak amplitude.

\subsection{Dynamical power spectra}\label{app-B1}

In order to quantify the main oscillation modes of the stellar motion and follow their evolution when initial conditions  vary, we construct a \emph{dynamical power spectrum}. For this, we Fourier analyse an orbit and plot the frequencies of the significant peaks of its power spectra as functions of the parameter describing a particular family of solutions. The smooth evolution of the frequencies is characteristic of regular motion, while the erratic spreading of the frequency values characterises the strongly chaotic behaviour of the system.

An example of a dynamical power spectrum is shown on the bottom panel in Fig.\,\ref{fig:map-bar-spectr}. In this case we analyse the oscillations of the radial (red) and azimuthal (black) coordinates of the objects along the rotation curve given by Eq.\,(\ref{eq:Vrot}), and plot their main frequencies as functions of their galactic distances  $R$. In the domains of regular motion, these frequencies (as well as their harmonics and possible linear combinations between them) evolve continuously when the value of $R$ is gradually varied. When the ILR and the 4/1 resonance are approached, the frequency evolution shows a discontinuity characterised by the erratic scatter of values when chaotic layers associated with separatrices are crossed. Inside the corotation and resonant islands, the frequencies split because of the qualitatively distinct dynamics which is intrinsic of the resonance.



\end{document}